\documentclass[fleqn,usenatbib]{mnras}

\usepackage{newtxtext,newtxmath}

\usepackage[T1]{fontenc}

\DeclareRobustCommand{\VAN}[3]{#2}
\let\VANthebibliography\thebibliography
\def\thebibliography{\DeclareRobustCommand{\VAN}[3]{##3}\VANthebibliography}

\usepackage{graphicx}	

\usepackage{amsmath}	

\usepackage{caption}
\usepackage{subcaption}
\usepackage{units}
\usepackage{academicons}
\usepackage{xcolor}
\definecolor{lime}{HTML}{A6CE39}

\usepackage{amsthm,enumitem}
\usepackage{float}

\newcommand{\tol}{\textrm{tol}}

\newcommand{\z}{\hat{\textbf{z}}}
\newcommand{\zcart}{\hat{\textbf{Z}}}
\newcommand{\zosc}{\textbf{z}}
\newcommand{\dz}{\delta \textbf{z}}

\newcommand{\q}{\hat{\textbf{q}}}
\newcommand{\qosc}{\textbf{q}}

\newcommand{\qoscddot}{\ddot{\textbf{q}}}

\newcommand{\dq}{\delta\textbf{q}}
\newcommand{\dqnorm}{\delta{q}}
\newcommand{\dqdot}{\delta\dot{\textbf{q}}}
\newcommand{\dqddot}{\delta\ddot{\textbf{q}}}
\newcommand{\dqdotnorm}{\delta\dot{q}}
\newcommand{\dqddotnorm}{\delta\ddot{q}}

\newcommand{\p}{\hat{\textbf{p}}}
\newcommand{\posc}{\textbf{p}}
\newcommand{\poscdot}{\dot{\textbf{p}}}
\newcommand{\dpp}{\delta \textbf{p}}
\newcommand{\dpdot}{\delta\dot{\textbf{p}}}

\newcommand{\qcart}{\hat{\textbf{Q}}}
\newcommand{\pcart}{\hat{\textbf{P}}}

\newcommand{\dt}{\textrm{d}t}

\newcommand{\cs}[1]{#1^{*}}

\title[Terrestrial Exoplanet Simulator]{Terrestrial Exoplanet Simulator (TES): an error optimal planetary systems integrator that permits close encounters.}

\author[P. Bartram and A. Wittig]{
Peter Bartram $^{1}$\thanks{E-mail: P.Bartram@soton.ac.uk} and Alexander Wittig$^{1}$
\\
$^{1}$University of Southampton, Southampton, United Kingdom
}

\date{Accepted XXX. Received YYY; in original form ZZZ}

\pubyear{2021}

\begin{document}
\label{firstpage}
\pagerange{\pageref{firstpage}--\pageref{lastpage}}
\maketitle

\begin{abstract}
We present TES, a new n-body integration code for the accurate and rapid propagation of planetary systems in the presence of close encounters. TES builds upon the classic Encke method and integrates only the perturbations to Keplerian trajectories to reduce both the error and runtime of simulations. Variable step size is used throughout to enable close encounters to be precisely handled. A suite of numerical improvements are presented that together make TES optimal in terms of energy error. Lower runtimes are found in all test problems considered when compared to direct integration using ias15. TES is freely available.

\end{abstract}




\section{Introduction}

Understanding and predicting the motion of the celestial bodies has been an active field of research since the times of Newton and Kepler and is still equally as important today. Few analytical solutions exist for planetary motion and scholars have instead turned to numerical n-body techniques. N-body problems range from the most general form found in the study of plasma and star cluster dynamics \citep{Aarseth1999} through to systems with more inherent structure such as protoplanetary disks \citep{Kokubo1996, Kokubo1998} or exoplanet systems \citep{Smith2009}. While integrators exist for the general case, by restricting one's self to systems that exhibit more structure one can leverage it to develop more efficient integration algorithms. In this paper we restrict ourselves to planetary systems whereby there exists a dominant central mass with any number of orbiting bodies. In this case, three key problems to address are:

\begin{enumerate}[leftmargin=*]
    \item Ensuring that solutions obtained remain accurate over the timescales required, in solar system formation and stability studies typically $10^9$ dynamical periods.
    \item Ensuring that simulations can be completed within the available computing time.
    \item Ensuring that integrators can precisely model close encounters between objects.
\end{enumerate}

IEEE 754 double precision floating-point numbers \citep{IEEE754}, typically used in scientific computing, have limited precision of about 16 decimal digits and as such errors are introduced when arithmetic operations are performed on them. All classes of numerical integrators are afflicted by this finite precision and over many time steps these \emph{round-off errors} can accumulate to corrupt the accuracy of solutions obtained. Furthermore, these are not the only source of error present as the choice of integrator and step size used can also allow for \emph{truncation error} to be introduced, while a less than careful implementation of intrinsic functions, such as the trigonometric functions, can lead to the introduction of \emph{bias error}. All of these error sources must be carefully considered when designing a numerical scheme to ensure solutions remain accurate for $10^9$ dynamical periods. 
When round-off error can be made to dominate it has been found that specific numerical techniques can be used to ensure that the distribution of errors are symmetrical. This has led to the creation of integration schemes that are optimal in the sense that they follow Brouwer's law \citep{Brouwer1937} and exhibit a growth in relative energy error over integrator time, $t$, proportional to $\sqrt{t}$. \citet{Rein2014}, with ias15, and \citet{Grazier2005} have both applied special numerical treatments to \emph{traditional integrators}, i.e. non-symplectic integrators, used in the solution of general ordinary differential equations (ODEs) that result in integrators that follow Brouwer's law and are therefore suitable for long-term integrations. Despite being optimal in the sense of Brouwer's law, these schemes are computationally expensive and require many, typically upwards of a thousand, evaluations of the force function per orbit. 

There are less computationally intensive means of ensuring invariants are preserved in long-term celestial mechanics integrations. For example, symplectic methods \citep{Forest1990, Kinoshita1990, Saha1992} can be used to place an upper bound on the truncation error of integrations by solving a system governed by a Hamiltonian that is slightly perturbed from that of reality. The use of symplectic methods ensures that the Poincaré invariants are conserved which in turn has favourable properties for energy and angular momentum conservation. \citet{Wisdom1991} created a symplectic mapping, known as the Wisdom-Holman (WH) map, that revolutionised the field by reducing computational times by an order of magnitude as compared to traditional integrators. The WH map has, and continues to be, the workhorse of the field and forms the basis of many diverse propagation tools \citep{Duncan1998, Chambers1999, Grimm2014, Rein2015}. In planetary systems, the WH map exploits the dominant contribution to the dynamics by the star and uses the fact that \emph{secondaries}, i.e. bodies in orbit around a more massive \emph{primary} body such as the Sun, move on perturbed Keplerian trajectories to split the system Hamiltonian into separate Keplerian and perturbation terms that can be solved independently within a time step. This splitting allows for the WH map to make only twenty evaluations of the force per orbit for typical applications. Obtaining the solution for the Keplerian term analytically requires the solution of Kepler's equation \citep{Battin1999}. There are many choices for solving Kepler's equation but universal variables \citep{Battin1999} are favoured for their versatility. Stumpff functions are typically used \citep{Danby1992} although other recent works shows that unbiased results can also be obtained without them \citep{Wisdom2015}.

One drawback of the WH mapping, and symplectic schemes in general, is that they must use a fixed timestep to ensure that symplecticity remains unbroken. Whilst not a problem if bodies remain well separated it means integrators are unable to handle close encounters which are typically defined as encounters between bodies within one Hill radius $r_H$. The ability to handle close encounters is highly important in celestial mechanics for modelling many problems. This includes: the threat of asteroids to the Earth \citep{Giorgini2008}, the behaviour of exoplanet systems after an instability event \citep{Rice2018, Bartram2021}, and the planet formation process itself \citep{Davies2014}.
Options exist that enable invariants to be conserved during close encounters when using the WH map \citep{Duncan1998, Chambers1999, Rein2019} but they fail to obtain the true trajectories of bodies during the encounter for the typical step sizes chosen. Therefore, to study realistic trajectories of bodies during close encounters traditional integrators such as Bulirsch-Stoer \citep{Bulirsch1966} or Everhart's RADAU \citep{Everhart1985} scheme are typically used.

In this work, we introduce our method, called the Terrestrial Exoplanet Simulator (TES), that aims to combine the accuracy and performance benefits of the analytical solution of the Keplerian motion, as found in symplectic schemes, with the flexibility of traditional integration schemes. We build upon the classic scheme of Encke, see e.g. \citep{Wiesel2010}, to create a perturbation method that can be integrated with a traditional integrator. Importantly, we show that in this framework it is possible for close encounters to be handled precisely and with a reduction in computational cost as compared to performing a \emph{direct integration} using the full n-body equation of motion. We show that through careful handling of round-off error through compensated summation \citep{Kahan1965, Higham1993} the Encke method can be made optimal in the sense that it follows Brouwer's law with a relative energy error comparable to that of ias15. We offer two implementations for TES, the \emph{standard configuration} where the implementation is purely in double precision floating point arithmetic and the \emph{extended configuration} where 80 bit extended precision, i.e long doubles, are used in the Kepler solver. The latter implementation enables a further reduction in error of an order of magnitude for the same computational cost when compared to using ias15. Both TES implementations are capable of accurately handling close encounters and TES has already been used to study exoplanet evolution in the presence of collisions \citep{Bartram2021}.

During the preparation of this manuscript, a similar method has been published by \citet{Hernandez2020}. Both methods have been developed independently and while they make use of similar concepts, the end result differs, with the most significant difference being that our method can be used to model close encounters.

We begin in Section~\ref{sec:TES Model} with a description of the components making up TES. Section~\ref{sec: DHEM} contains the derivation of the equations of motion used. The solution of the analytical part of our model is described in Section~\ref{sec: analytical solution} and the numerical part in Section~\ref{sec: numerical solution}. Section~\ref{sec: implementation details} contains details about specific numerical techniques implemented to ensure TES follows Brouwer's law. We show the impact of each of these techniques in Section~\ref{sec: validation of implementation}. Beginning in Section~\ref{sec: numerical experiments}, the second half of this paper contains a series of numerical experiments. In particular, Section~\ref{sec: numerical experiment longterm integrations} contains long-term integrations and Section~\ref{sec: numerical experiments apophis} shows the performance in the presence of close encounters. We offer some concluding remarks in Section~\ref{sec: conclusions}.

\section{TES Model}
In this section we begin by giving an overview of the method and then describe in detail the mathematical model, coordinate system, and force function used in TES.

\label{sec:TES Model}
\subsection{General Encke Method}

\begin{figure}
    \centering
    \includegraphics[width=0.475\textwidth]{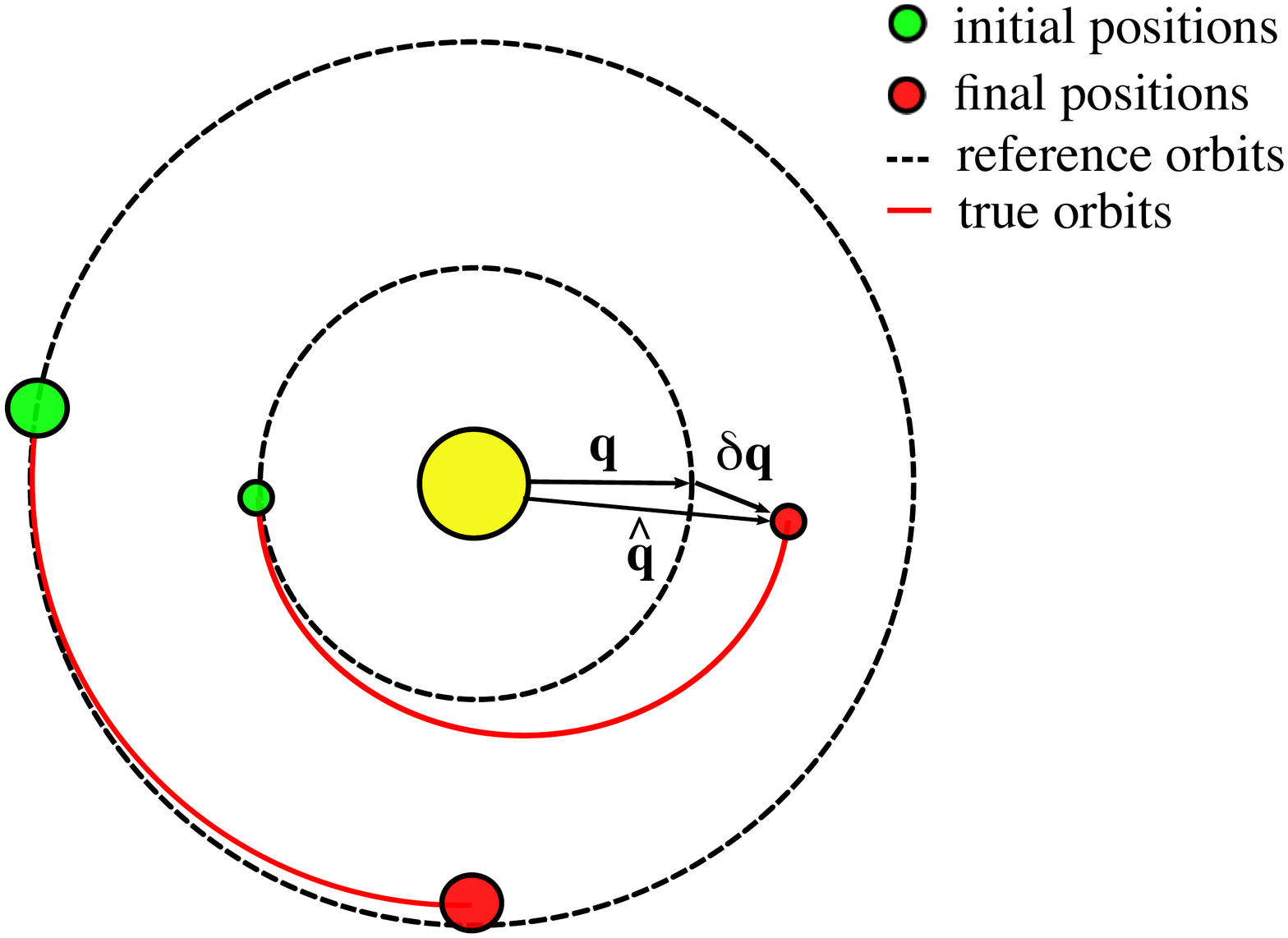}
    \caption{A three-body Encke method. For the inner planet, the position on the reference orbit, $\qosc$, is shown along with the perturbation from it, $\dq$. The position on the true orbit, $\q$, is also shown. Deviations from the reference orbits are greatly exaggerated for clarity.}
    \label{fig:encke orbital diagram}
\end{figure}

In the Encke method, the position of a given body $\q$ is made up of two terms: 

\begin{enumerate}[leftmargin=*]
\item $\qosc$, the two-body reference trajectory,
\item $\dq$, the perturbation to this two-body motion,
\end{enumerate}

such that $\q = \qosc + \dq$. Figure \ref{fig:encke orbital diagram} illustrates this concept, where the reference trajectory, $\qosc$ , can be obtained analytically at any future time by solving Kepler's equation and applying the so called $f$ and $g$ functions \citep[p.~156]{Battin1999}. In contrast, the perturbation term, $\dq$, must be obtained through numerical integration. The advantages of the Encke method over, e.g., a full n-body integration stem from the fact that the ratio of  $\nicefrac{|\dq|}{|\qosc|}$, henceforth called the \emph{delta ratio}, is much smaller than unity meaning that a given absolute precision in $\q$ can be obtained with lower relative precision in $\dq$ which means less computation by the numerical integrator. In order to keep the delta ratio small one occasionally needs to update the reference trajectory to the current true orbit, a process called \emph{rectification}.
With a single rectification per orbit a delta ratio of $10^{-2}$ can reasonably be expected for simulations of the outer solar system whereas for the inner solar system this ratio remains below $10^{-4}$. With such delta ratios maintained, the analytical solution of the reference trajectory becomes the primary source of numerical error. In order for this method to work, it is crucial that this propagation is precise, in the case of this work this means down to machine precision.

We continue by introducing a general form for creating an Encke method through two arbitrary governing Hamiltonians which therefore requires that we are integrating a conservative system without any dissipative effects present. Later, in Section~\ref{sec: DHEM}, we use this form to derive our method for two specific Hamiltonians in our chosen coordinate system.
We choose to work in canonical coordinates, and throughout this work the true state vector is denoted by $$\z = (\q, \p)^{T}$$ where $\q$ and $\p$ are conjugate position and momentum vectors. Similarly, the reference orbit state vector is $$\zosc = (\qosc, \posc)^{T}$$ where $\qosc$ and $\posc$ represent the position and momentum components of the reference orbits. Both $\z$ and $\zosc$  are assumed to be in the same coordinate system. Therefore, the Encke method is
\begin{equation}
    \nonumber
    \z = \zosc + \dz
\end{equation}
and the perturbation term, henceforth called simply \emph{the deltas}, for which we need to derive the equations of motion, is
\begin{equation}
    \dz = \z - \zosc.
    \label{eq: deltas}
\end{equation}

The time evolution of $\z$ and $\zosc$ for any conservative system is governed by their respective Hamiltonians $\hat{\mathcal{H}}(\z)$ and $\mathcal{H}(\zosc)$. Making use of the canonical structure matrix 
 \begin{equation}
 \nonumber
J \equiv
  \begin{bmatrix}
    0 & +I \\
    -I & 0
  \end{bmatrix},
  \label{eq: canonical structure matrix}
\end{equation}
where $I$ is the identity matrix of appropriate dimensions for a given problem, one can apply Hamilton's equations to yield the equations of motion as
\begin{equation}
\dfrac{\textrm{d}}{\textrm{d}t} \z = J \nabla_{\z} \hat{\mathcal{H}}(\z),
\quad
\dfrac{\textrm{d}}{\textrm{d}t} \zosc = J \nabla_{\zosc} \mathcal{H}(\zosc).
\label{eq: deltas derivatives}
\end{equation}

Taking the time derivative of Eq.~\eqref{eq: deltas} and replacing appropriate terms with those from Eq.~\eqref{eq: deltas derivatives} gives a formula for finding the equations of motion for the deltas themselves as
\begin{equation}
\dfrac{\textrm{d}}{\textrm{d}t} \dz = \dfrac{\textrm{d}}{\textrm{d}t} \left(\z - \zosc \right) = 
J \left(  \nabla_{\z} \hat{\mathcal{H}}(\z) -  \nabla_{\zosc} \mathcal{H}(\zosc) \right).
\label{eq: encke method vector form}
\end{equation}

\subsection{Encke Method: Democratic Heliocentric (ENCODE)}
\label{sec: DHEM}
In Cartesian coordinates, we define the state vector for this system as $$ \hat{\textbf{Z}}(t) = (\hat{\textbf{Q}}(t), \hat{\textbf{P}}(t))^{T}$$ where $\hat{\textbf{Q}}_i(t)$ is the generalised position vector and $\hat{\textbf{P}}_i(t)$ is the momentum vector conjugate to it.
If the bodies are only interacting through mutual Newtonian gravity then the time evolution of the system in Cartesian coordinates is governed by the gravitational n-body Hamiltonian \citep{Liemkuhler2005}
\begin{equation}
\hat{\mathcal{H}}(\zcart) = \sum^n_{i=0} \dfrac{|\pcart_i|^2}{2\,m_i} - 
\sum^{n}_{i=0} \sum^n_{j = i+1} \dfrac{G m_i \, m_j}{|\qcart_j - \qcart_i|}
\label{eq:nbody cartesian}
\end{equation}
where the subscript $i$ refers to the $\textrm{i}_{th}$  body in the system. Throughout this work $i=0$ is reserved to refer to the central body.

There are no obvious Hamiltonian splittings of Eq. \eqref{eq:nbody cartesian} that allow for the dominant Keplerian motion of the secondary bodies about the primary, due to the dominant central mass, to be isolated from the general evolution of the system. For this it is necessary to introduce a different coordinate system. There are several coordinate systems available and \citet{Hernandez2017} give a good overview of the canonical coordinate systems for the n-body problem. The Jacobi coordinate system was used by \citet{Roy1988} to create an Encke method to simulate the outer planets of our solar system. In this coordinate system, each body has a reference orbit taken with respect to a different moving centre of mass that depends on the position and mass of all other bodies who's orbits are smaller than its own. Therefore, as the relative size of the orbits of bodies changes in a system it becomes necessary to recalculate reference trajectories with respect to a new moving centre of mass. To avoid this complication and the numerical error it could introduce we instead opt to use democratic heliocentric (DH) coordinates \citep{Duncan1998} which are common in celestial mechanics \citep{Chambers1999, Grimm2014}.
In DH coordinates the equations of motion are such that the position of each body is expressed relative to the central body, which we denote with the index zero throughout this work. The momentum of each body, however, is expressed relative to the barycentre of the system. The coordinate change from Cartesian to democratic heliocentric coordinates is given by
\begin{equation}
\q_i = 
\begin{cases}
\qcart_i - \qcart_0, & \textrm{if} \, i \neq 0 \\
 \dfrac{1}{M} \sum^n_{j=0} m_j \qcart_j, & \textrm{if} \, i = 0
 \end{cases}
 \label{eq: coordinate transform position}
\end{equation}
\begin{equation}
\p_i = 
\begin{cases}
\pcart_i - \dfrac{m_i}{M} \sum^n_{j=0} \pcart_j & \textrm{if} \, i \neq 0 \\
\sum^n_{j=0} \pcart_j & \textrm{if} \, i = 0.
\end{cases}
\label{eq: coordinate transform momenta}
\end{equation}
where M is the total system mass, i.e. $M = \sum^n_{j=0} m_j$. Therefore, $\q_0$ and and $\p_0$ are the centre of mass and momentum of the system, respectively.
The Hamiltonian, $\hat{\mathcal{H}}(\z)$, as a function of the previously defined state vector, $\z$, in these new coordinates is
\begin{equation}
\hat{\mathcal{H}}(\z) = \hat{\mathcal{H}}_{star} + \hat{\mathcal{H}}_{kep} + \hat{\mathcal{H}}_{pert}
\label{eq: hamiltonians split up}
\end{equation}
where
\begin{align}
\nonumber
\hat{\mathcal{H}}_{star} =& \dfrac{1}{2 m_{0}} \sum_{i=1}^n |\p_i|^2, \\
\nonumber
\hat{\mathcal{H}}_{kep} =& \sum_{i=1}^n \left(  \dfrac{|\p_i|^2}{2\,m_i} - \dfrac{G\,m_i\,m_0}{|\q_i|}   \right), \\
\nonumber
\hat{\mathcal{H}}_{pert} =& \sum_{i=1}^n \sum_{j=i+1}^n  \dfrac{G\,m_i\,m_j}{|\q_j - \q_i|}. 
\end{align}

Each of the three components of $\hat{\mathcal{H}}$ are physically identifiable:
\begin{enumerate}[leftmargin=*]
\item  $\hat{\mathcal{H}}_{star}$ is the motion of the central body,
\item  $\hat{\mathcal{H}}_{kep}$ is the Keplerian motion of the secondary bodies about the central body,
\item $\hat{\mathcal{H}}_{pert}$ is the gravitational interactions between secondary bodies. 
\end{enumerate}

An additional fourth term that only depends upon $\p_0$ and represents the motion of the centre of mass has been excluded under the assumption of a stationary barycentre, and it is therefore unnecessary to propagate $\q_0$ and $\p_0$. 

Note that Equation \eqref{eq:nbody cartesian} can also be used to obtain a Hamiltonian describing a system of particles that only interact with the central body by enforcing that $j=0$ thereby removing the gravitational interactions between secondaries.
After applying the same coordinate transformation from Eqs.~\eqref{eq: coordinate transform position} and \eqref{eq: coordinate transform momenta} the reference orbit Hamiltonian, $\mathcal{H}(\zosc)$, as a function of the previously defined state vector, $\zosc$, in democratic heliocentric coordinates reads
\begin{equation}
\nonumber
\mathcal{H}(\zosc) = \sum_{i=1}^n \left( \dfrac{|\posc_i|^2}{2\,m_i} - \dfrac{G \, m_0 \, m_i}{|\qosc_i|} \right). \\
\end{equation}

Inserting $\hat{\mathcal{H}}(\z)$ and $\mathcal{H}(\zosc)$ into the general Encke form in Eq.~\eqref{eq: encke method vector form} yields the equations of motion for the deltas in democratic heliocentric coordinates as
\begin{align}
\dqdot_i = \dfrac{\dpp_i}{m_i} + \dfrac{1}{m_0} \sum^n_{j=1} \left( \posc_j + \dpp_j \right),
\label{eq: delta q dot} \\
\dqddot_i = \dfrac{\dpdot_i}{m_i} + \dfrac{1}{m_0} \sum^n_{j=1}\left( \poscdot_j + \dpdot_j  \right),
\label{eq: delta q ddot}
\end{align}
\vspace{-3mm}
\begin{equation}
\begin{split}
\dpdot_i =&\, G m_i\,m_0 \left(\dfrac{\left( \qosc_i + \dq_i \right)}{|\qosc_i + \dq_i|^3}  -\dfrac{\qosc_i}{|\qosc_i|^3} \right) + \\ &\sum_{j=1}^n  G\,m_i\,m_j\, \dfrac{\left(\qosc_j + \dq_j \right)-\left(\qosc_i + \dq_i \right)}{\left|\left(\qosc_j + \dq_j \right) - \left(\qosc_i + \dq_i \right)\right|^3}.
\label{eq: delta p dot}
\end{split}
\end{equation}

Here, and throughout this work, dots are used to signify the time derivative of a variable. For reasons discussed later in Section~\ref{sec: numerical solution} we take an additional time derivative of $\dqdot$ to obtain $\dqddot$. There are several key features to note about these equations. The summation in the trailing term in Eq.~\eqref{eq: delta p dot} starts at an index of $1$ meaning that it only captures interactions between secondary bodies; the gravitational interaction with the central body, with an index of $0$, is captured by the leading term instead. In this term, it can be seen that there is cancellation between similar terms that depends upon only the size of $\dq_i$. Therefore, so long as this term remains small then $\dpdot_i$ will generally also remain small in comparison to the the reference trajectory derivative, $\poscdot$. The exception to this are close encounters between secondaries, where this term can grow significantly. Again, it is this reduction in the relative size of the term to be integrated that leads to the performance increases from an Encke method.

Typically, the Encke method as used in astrodynamics assumes massless secondaries, and, as such, that the centre of mass of the central body is coincident with the barycentre, i.e., the location to which reference orbits are taken is fixed. This is not the case in celestial mechanics, however, due to the relatively high mass ratio of planets in comparison to their host stars, a ratio of approximately $10^{-3}$ for our solar system. 
As an example, consider a two-body problem consisting of Jupiter and the Sun. In this case, the elliptical motion of the Sun about the Sun-Jupiter barycentre is captured by the trailing term in Eq.~\eqref{eq: delta q dot} as the negative momentum of the star
$$\p_{star} = -\frac{1}{m_0} \sum^n_{j=1} \left( \posc_j + \dpp_j \right).$$
This shows that even in a purely two-body case there will still be a deviation from the reference trajectory around the Sun over time, and the magnitude of this deviation is inversely proportional to the ratio of the mass of the star $m_0$ to the mass of the other bodies within the system $m_p \equiv \sum_{j=1}^n m_j$. We call this ratio the \emph{system mass ratio} and define it as $\frac{m_p}{m_0}$. Practically, this means that 
systems with a small system mass ratio have the greatest potential performance gains.
Alternatively, in systems with many bodies an axially symmetric distribution of mass reduces the motion of the central body through a cancellation of terms in $p_{star}$. Clearly, a distribution such as this is non-physical for planetary systems, but typical accretion disks in planetary formation are symmetric enough to see cancellation.

We point out that for the previously discussed reasons it is no longer necessary to evolve the motion of the central body as an separate set of equations thereby reducing $n$ by $1$. However, whereas a pure n-body integration can take advantage of a second order formulation of the equations of motion to reduce the number of equations to be integrated, especially in the case where the motion is not velocity dependent, this is no longer the case for these equations where two first order ODEs, Eqs.~\eqref{eq: delta q dot} and \eqref{eq: delta p dot}, must instead be integrated. Note that this does not double the computational cost. The additional cost in the RHS is small and scales only as $\mathcal{O}(n)$ and therefore the dominant $\mathcal{O}(n^2)$ interaction term remains unchanged. Furthermore, the integration procedure itself performs identical operations for each first order ODE meaning that vectorisation, either manual or automatically performed by the compiler, can recover some of the additional computational cost. Finally, this does not have a noticeable performance penalty when calculating the reference trajectories.

\subsection{Analytical Solution}
\label{sec: analytical solution}

In this section we describe how we solve the two-body problem with universal variables making use of the $f$ and $g$ functions \citep{Danby1992}. If we assume that the integration of the perturbation terms can be performed in a way that ensures the truncation error is below floating point precision for $\dz$ then the overall precision of the scheme depends upon the precision in the solution of the Keplerian motion. As such, we require a highly accurate solver implementation that is also non-biased to ensure that the energy growth follows Brouwer's law for long duration integrations. Several modern implementations of Kepler solvers exist that are suitably accurate, e.g. \citep{Wisdom2015, Rein2015}. We have chosen to follow the implementation presented in WHFAST \citep{Rein2015} but with some additional numerical improvements specific to our needs, these are discussed in Section~\ref{sec: implementation analytical}.

We wish to solve $n-1$ independent two-body problems for which the time evolution is governed by $\hat{\mathcal{H}}_{star}$ in Eq.~\eqref{eq: hamiltonians split up}. Typically, the f and g functions use a reduced mass parameter such that $\mu = G \left(m_0+m_i\right)$; however, in democratic heliocentric coordinates, used here, the reduced mass is given by $\mu = G m_0$ \citep{Grimm2014}.
In the following discussion we focus on a single two-body problem from the vectors $\qosc$ and $\posc$ above. For the remainder of this section $\qosc$ and $\posc$ refer to the position and momentum of a single body undergoing Keplerian motion. As a result, we drop the index referring to each body: an index of zero now refers to values at the start of an integration step. After using the WHFAST algorithm to solve for the universal anomaly and obtain the $G$-functions \citep{Rein2015}, the $f$ and $g$ functions used are 
\begin{align}
    f = - \dfrac{\mu G_2}{|\qosc_0|},
    \quad
    \nonumber
    \dot{f} = - \dfrac{\mu G_1}{|\qosc_0| |\qosc|}, \\
    g = \textrm{d}t - \mu G_3,
    \quad
    \dot{g} = - \dfrac{\mu G_2}{|\qosc|}
    \label{eq: g}
\end{align}
where $\textrm{d}t$ is the time step. This allows for the solution to the Kepler problem, after a timestep of $\textrm{d}t$, to be obtained from the initial values of position and momenta and a linear combination of them to be applied to each as an update term, $\Delta \qosc$ and $\Delta \posc$,  such that 
\begin{align}
    \qosc =& \qosc_0 + \Delta \qosc = \qosc_0 + \left(f \qosc_0 + g \dfrac{\posc_0}{m}\right),
    \label{eq: update qosc} \\
    \posc =& \posc_0 + \Delta \posc = \posc_0 + m \left(\dot{f} \qosc_0 + \dot{g} \dfrac{\posc_0}{m}\right).
    \label{eq: update posc}
\end{align}

When formulated in this manner, the smaller $\textrm{d}t$, relative to the orbital period, the smaller the size of the bracketed terms and therefore summing them first is more numerically robust. Additionally, TES uses a value of $\textrm{d}t$ approximately $\nicefrac{1}{300}^{\mathrm{th}}$ of an orbit, roughly an order of magnitude smaller than, e.g., WHFAST and this means that the relative sizes of $\Delta \qosc$ to $\qosc_0$ and $\Delta \posc$ to ${\posc_0}$ enables compensated summation to be used to further reduce round-off error; our algorithm for this is described in Section~\ref{sec: implementation rectification}.

\subsection{Numerical Solution}
\label{sec: numerical solution}
While there are many numerical integrators to choose from, a particularly efficient and accurate choice for astrophysical problems is the RADAU scheme of \citet{Everhart1974}.  A useful feature of RADAU is that once a polynomial is fitted to the force, in the manner discussed below, it is possible to integrate it analytically one or multiple times allowing for both, e.g., velocity and position to be obtained from the same polynomial. \citet{Everhart1985} found that his scheme is best suited for directly integrating second order ODEs, e.g. $\ddot{y} = F(y, t)$, without reduction to a pair of first order equations. In fact, for the same number of evaluations of the force function he found the solution to second order ODEs could be as much $10^{6}$ times more precise than if they were reduced to first order and integrated.
In the derivation of our equations of motion we have had to reduce the system to a pair of first order equations, and integrating $\dqdot$ and $\dpdot$ in this manner does in fact cause a reduction in performance. We observed that integrating $\dqddot$ and $\dpdot$ is enough to circumvent this problem and opt to integrate these equations in TES. As discussed in Section~\ref{sec: DHEM}, the effect of this choice on the computational cost is minor.

\citet{Rein2014} further refined the RADAU method with two new algorithms, one for convergence criteria and one for stepsize control, and additionally included compensated summation to ensure a symmetrical distribution of round-off errors. Their new implementation is $15^{\mathrm{th}}$ order and is called ias15. This is the version we have chosen to base our implementation on to integrate $\dqddot$ and $\dpdot$. Next, we introduce just enough of the ias15 algorithm to discuss our modifications to it. Only the process of integrating $\dqddot$ is discussed but a similar process is also followed for $\dpdot$.

We need to simultaneously solve $3n$ equations of the form $$ \dqddotnorm = F(\qosc, \posc, \dq, \dpp, t),$$ one for each directional component of $n$ bodies. In order to do this, RADAU expands the acceleration, $\dqddotnorm$,  in time, $t$, into a truncated series such that
\begin{equation}
    \dqddotnorm(t) \approx \dqddotnorm_0 + b_0 h + b_1 h^2 + ... + b_6 h^7
    \label{eq: series ias15}
\end{equation}
where $h = t / \textrm{d}t$, $\textrm{d}t$ is the size of an integration step, and $t_0$ and $\dqddotnorm_0$ are the time and acceleration at the start of an integration step. The $b$ coefficients are fitted though an iterative predictor-corrector process that performs an evaluation of $\dqddotnorm$ at the start of an integration step, $t=t_0$, and at seven \emph{substeps} within the integration step. The sampling locations in time for each substep, $c_i$ where $i=1...7$, are chosen in accordance with Radau quadrature spacings \citep{Radau1880} to maximise the order of the scheme. Once the coefficients, $b_i$, are obtained to a sufficient precision then Eq.~\eqref{eq: series ias15} can be integrated analytically to obtain an estimate of $\dq$ at the end of a step, $t_1 = t_0 + \dt$, as
\begin{equation}
\begin{split}
    \dqnorm(t_1) &\approx  \dqnorm_0 + \dt \; \dqdotnorm_0 \; + \\
    &\dt^2 \left(\dfrac{\dqddotnorm_0}{2} + \dfrac{b_0}{6} + \dfrac{b_1}{12} + \dfrac{b_2}{20} + \dfrac{b_3}{30} + \dfrac{b_4}{42} + \dfrac{b_5}{56} + \dfrac{b_6}{72}  \right).
    \nonumber
\end{split}
\end{equation}
At the start of a step an analytical continuation of the curve fitted to $\dqddotnorm$ via the accurate $b_i$ values calculated during the previous step are used to generate a predictor in the form of the values of $b_i$ to use in the current step.
To ensure convergence when iterating to obtain the coefficients $b_i$ a \emph{convergence criterion} is required. We opt to use a convergence criterion similar to that in ias15 and therefore monitor the change in the final coefficient in our truncated series, i.e $b_6$, from one iteration to the next, we call this change $\Delta \textbf{b}_6$ which is a vector containing coefficients for all $3n$ equations. We then compare the maximum change to the maximum magnitude of the reference orbit acceleration, $\qoscddot$, and we terminate when
\begin{equation}
\dfrac{\left\| \Delta \textbf{b}_6 \right\|_\infty}{\left\| \qoscddot_0 \right\|_\infty} < 10^{-16}.
\label{eq: convergence criteria}
\end{equation}

We find that this criterion performs better in our use case than if $\dqddot$ is used in the place of $\qoscddot$ in Eq.~\eqref{eq: convergence criteria} which is more typical. The typical criterion would ensure that the change in the coefficients in the series expansion of the acceleration of the deltas are precise to floating point precision. By design, in TES, the deltas are much smaller than the reference orbit terms and therefore it is unnecessary to converge this far to achieve a combined relative tolerance in $\q+\dq$ of $10^{-16}$ for use within an integration step, e.g. in the predictors.

In contrast to the convergence criterion, as we wish to suppress truncation error across long duration integrations, it is a requirement that the step size, $\dt$, is chosen such that it controls the truncation error in the series expansion, Eq.~\eqref{eq: series ias15}, itself. We monitor the truncation error for all $3n$ equations which is estimated by $\epsilon$. To do this, we monitor the absolute value of the $b_6$ coefficients for all $3n$ equations, which we call $\mathbf{b}_6$, relative to the acceleration of the deltas at the start of an integration step to obtain an estimate of the smoothness of the acceleration of the deltas over a given step as
\begin{equation}
\epsilon = \dfrac{\left\| \textbf{b}_6 \right\|_\infty}{\left\| \dqddot_0 \right\|_\infty}.
\label{eq: error estimate}
\end{equation}
The value of $\epsilon$ is then used to determine the next step size, $\dt_{n+1}$, as
\begin{equation}
    \dt_{n+1} = \dt \left( \dfrac{\tol}{\epsilon} \right)^{1/7}
    \label{eq: dt n plus one}
\end{equation}
where $\tol$ is a dimensionless tolerance parameter. We offer no analytical reasoning for choosing a value of $\tol$; however, numerical experiments demonstrating its effect are shown in Section~\ref{sec: numerical experiments}. We find that a default value of $10^{-6}$ is suitable for maintaining Brouwer's law for $10^9$ dynamical periods for simulations of the inner solar system and for handling close encounters. We have not included a step rejection algorithm as we found little benefit.

\subsection{Rectification}
Rectification is the name given to the process whereby a new reference trajectory is taken by adding the current reference trajectory together with the deltas. Our algorithm for doing this can be found in Section~\ref{sec: implementation rectification}. Rectification is important and the frequency with which is it performed has two conflicting effects on the efficiency of the scheme that must be balanced. Firstly, because rectification causes the deltas to be set to zero the analytical continuation used to obtain a prediction of the values of $b_i$ becomes much less precise in the integrator during the subsequent step and therefore increases the number of iterations required for convergence, typically by one. Secondly, in contrast, rectifying causes the size of the deltas to be reduced to zero and therefore reduces the computational cost on the integrator in many subsequent steps. We experimented with several rectification schemes based upon the size of the deltas relative to the reference trajectories but found the optimal cutoff value depends largely upon the mass ratio of the system. Ultimately, we found that a scheme where a rectification for all particles is performed between once and twice per orbit of the shortest period body is simple to implement and provides a good balance between the two effects. We choose to rectify according to the golden ratio and perform $1.618...$ rectifications per orbit of the body with the smallest period in the system to help avoid any possible bias effects due to resonances. We combine this with a fall back method whereby we also rectify if the delta ratio exceeds a given value, typically $10^{-3}$.

\section{Implementation Details}
\label{sec: implementation details}

\begin{figure}
    \centering
    \includegraphics[width=0.475\textwidth]{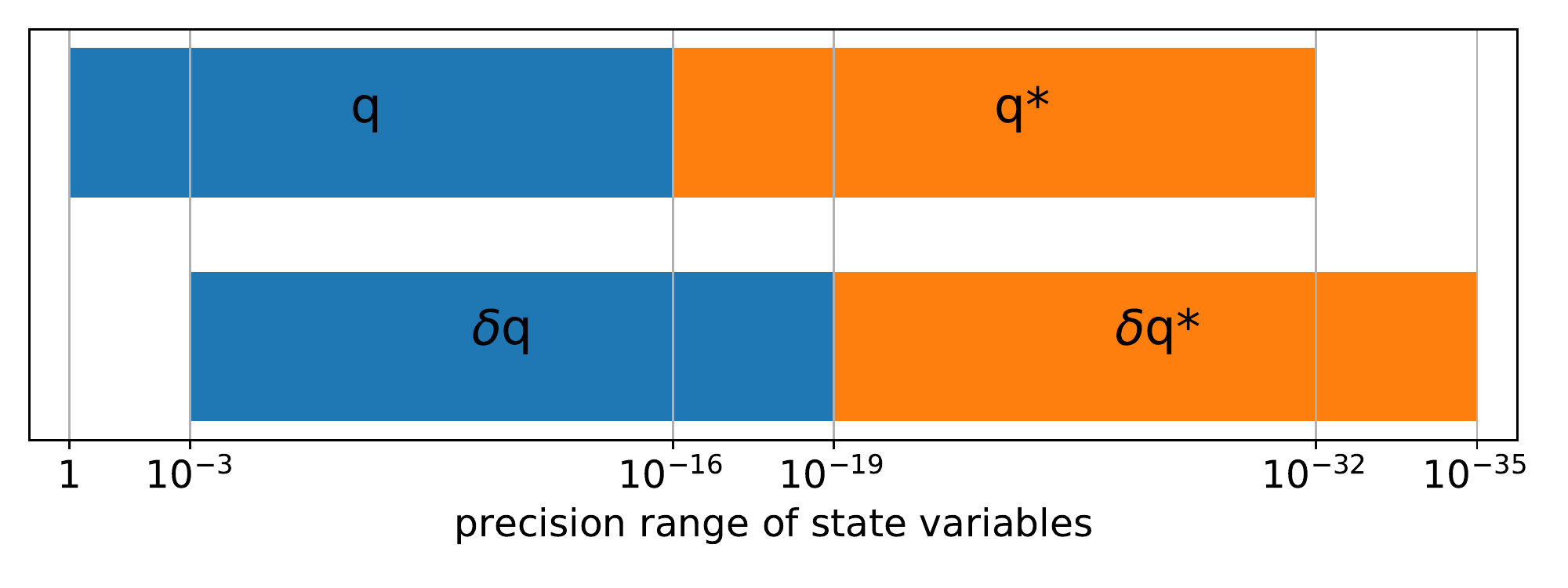}
    \caption{Precision ranges of terms within TES using an example where $q$ is of unity magnitude. The blue area shows a double precision floating point variable and the orange area shows the range covered by the associated compensation variable. The cross-hatched area indicates the key region of precision where extended precision floating point arithmetic can be used to improve the overall performance of TES.}
    \label{fig:absolute state variable precision}
\end{figure}

\begin{figure}
    \centering
    \includegraphics[width=0.475\textwidth]{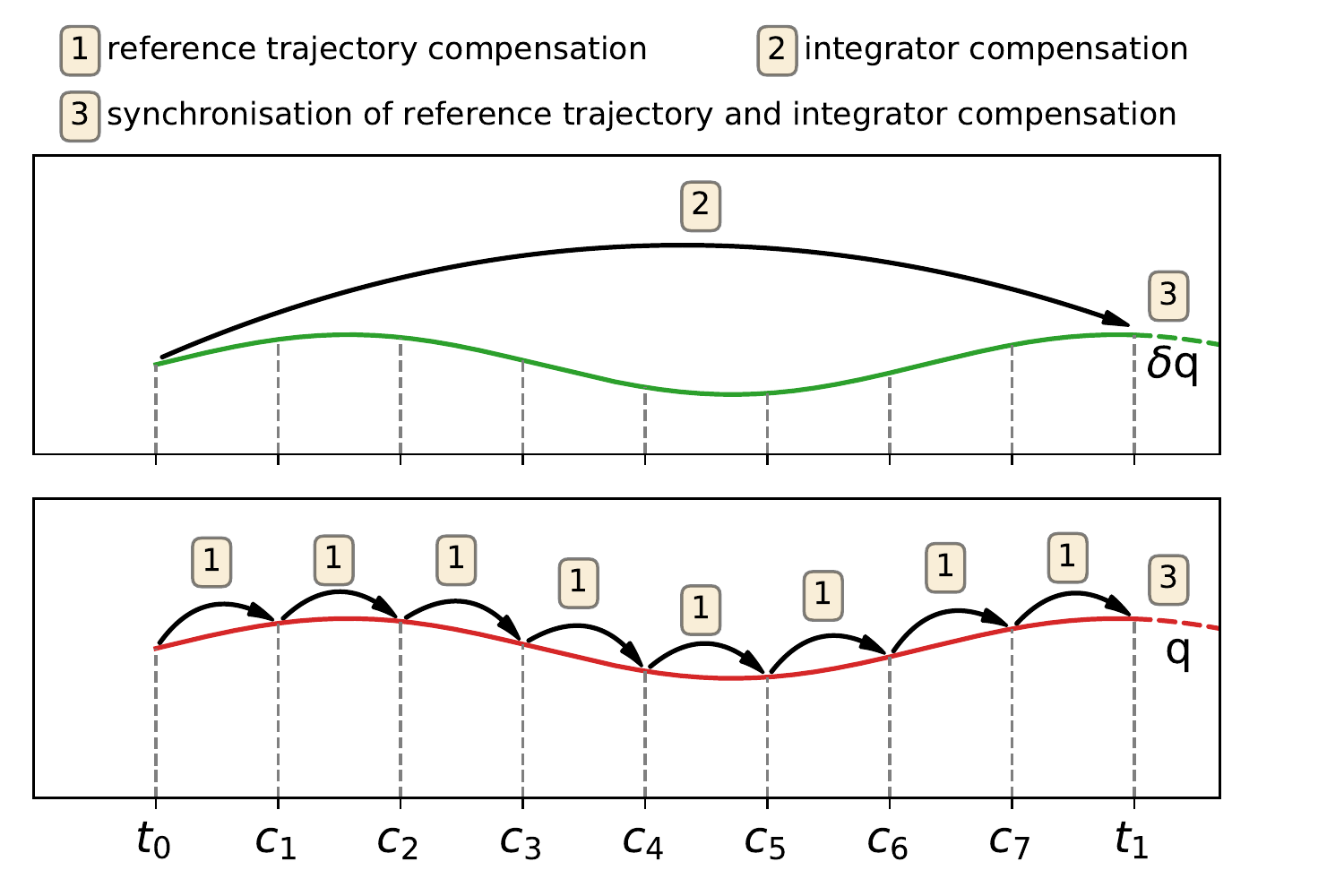}
    \caption{Locations in a single step of the RADAU integrator, beginning at $t_0$ and ending at $t_1$, where compensated summation is applied. Each value of $c_i$ is an integrator substep location at which a reference trajectory must also be calculated. The bottom panel shows the calculation of the reference trajectories where compensated summation is used at each forward step of the Kepler solver, marked by the label $1$, to maximise precision. The top panel shows the calculation of the deltas in the integrator. Here, a compensation variable is used to keep track of lost precision across an entire integration step, as shown by label 2. Finally, at the end of the integration step, $t_1$, label 3, compensated summation is used to combine the separate compensation terms. Compensated summation is also used to reduce error during rectification but this is not shown here.}   
    \label{fig: compensation diagram}
\end{figure}

In this section we introduce a collection of numerical implementation features that improve the overall energy conservation of TES for long integration times and enable it to handle close encounters. A key component used through this work is compensated summation \citep{Kahan1965}. Compensated summation allows for the error made during the summation of two double precision floating point numbers to be obtained and kept as an additional double precision number, known as the compensation variable. The error made is then subtracted from any future additions. This is applicable, e.g., when performing the final update of the state vector in an integration. We use this technique to ensure a symmetrical distribution of round-off error and minimise the total energy error in simulations. Compensated summation requires an additional \emph{compensation variable} be kept for each variable being compensated to keep track of lost precision. To this end, we define a composite datatype, written as $\{\Box, \cs{\Box} \}$ and comprised of two components: the variable itself, and its compensation variable which is denoted with a star; each component is stored in the computer as a double precision floating-point number. In infinite precision, the true value represented by a composite datatype $\Sigma = \Box - \cs{\Box}$ where the minus sign is due to the Kahan summation algorithm.  Figure~\ref{fig:absolute state variable precision} shows the relative ranges in magnitude that are covered by the variables used in TES. Compensated addition and subtraction operations are denoted by $\oplus$ and $\ominus$ respectively. We find it advantageous to use compensated summation in three ways in addition to internally within the integrator:

\begin{enumerate}[leftmargin=*]
    \item Propagating the reference trajectories.
    \item Combining the reference trajectories and deltas.
    \item Ensuring precision is maintained across rectifications.    
\end{enumerate}

\subsection{Encke Method: Democratic Heliocentric (ENCODE)}
\label{sec: implementation DHEM}

Equation~\eqref{eq: delta p dot} contains a subtraction between two terms of similar size and, owing to the finite precision of floating point numbers, this causes cancellation of significant digits which leads to a large decrease in the relative precision of $\dpdot$. The loss of relative precision here depends on how small the difference between the terms is. In particular, after a rectification the difference can be very small, e.g. $10^{-12}$ has been observed. In addition to the risk of introducing numerical error when calculating the acceleration this also poses a problem for the step size control algorithm in Eq.~\eqref{eq: error estimate} and \eqref{eq: dt n plus one}. The algorithm works by ensuring that the step size is chosen such that the acceleration approximated by the expansion in Eq.~\eqref{eq: series ias15} is smooth to  a precision of $\tol$. Oftentimes, the numerical cancellation in Eq.~\eqref{eq: delta p dot} means that the required degree of smoothness cannot be met. This can lead to a situation where the step size will shrink uncontrollably as the algorithm tries to shrink the step size further to reach an unattainable smoothness. To remedy this situation, it is possible to reformulate the problematic term to avoid the subtraction of like terms and rewrite the term $\dpdot$ as \citep[p.~449]{Battin1999}
\begin{align}
\nonumber
&u = \dfrac{\dq \cdot \left( \dq - 2\q \right)}{\q \cdot \q}, \\
\nonumber
&v = \dfrac{-u \left( 3 + 3u + u^2 \right)}{1 + (1+u)^{\frac{3}{2}}}, \\
&\dpdot_i =\, \dfrac{G m_i\,m_0}{|\q_i|^3}  \left(v \q - \dq \right) +
\sum_{j=1}^n  G\,m_i\,m_j\, \dfrac{\q_j  -\q_i }{|\q_j  - \q_i|^3}.
\nonumber
\end{align}
where $\dpdot$ is now obtained without loss of significance. This is the equation that we have implemented and it decreases numerical error as well as preventing a step size lockup from occurring.

\subsection{Analytical Solution}
\label{sec: implementation analytical}
Due to the relative size of the reference trajectories $\qosc$ compared to the deltas $\delta \qosc$, the dominant contribution to error growth stems from roundoff errors in $\qosc$. To minimise this potential we have used compensated summation in the final update step of the $f$ and $g$ functions, Eq.~\eqref{eq: update qosc} and \eqref{eq: update posc}. With compensated summation they become
\begin{equation}
\begin{split}
    \{\qosc, \qosc* \} = \{\qosc_0,\qosc_0*\} \oplus \Delta \qosc, \\ 
    \{\posc, \posc* \} = \{\posc_0,\posc_0*\} \oplus \Delta \posc.
    \label{eq: update qosc compensated}
\end{split}
\end{equation}
Due to the relatively small update terms, $\Delta \qosc$ and $\Delta \posc$, this allows for the value of $\qosc$ and $\posc$ to be maintained to machine precision across Kepler solver steps. 
Owing to non-linearity when calculating the $G$-functions it is more precise to take $k$ steps of size $\tau$ than one step of size $k \tau$. We therefore go further and also minimise the value of $\dt$ in Eq.~\eqref{eq: g} (the simulation time passed since the $f$ and $g$ function basis vectors were calculated) which in turn decreases the size of $\Delta \qosc$ and $\Delta \posc$. To do this, we only ever take a single step in the Kepler solver before we calculate new basis vectors, i.e., we calculate new basis vectors at the start of each step as well as at each substep required by the integrator. The locations in time that we perform both the compensation in Eq.~\eqref{eq: update qosc compensated} and a recalculation of basis vectors are illustrated in Fig.~\ref{fig: compensation diagram} and are marked by the label $1$. Here, it is shown that the universal variables compensation is used at the start of a step, $t_0$, at the end of a step $t_1$, and also at all substeps required by the integrator, $c_i$.

Whilst the standard TES configuration only makes use of double precision floating-point arithmetic throughout, there is also a build configuration that allows for the selected use of extended precision arithmetic through the \emph{C long double} datatype. We only use extended precision to perform the entirety of the reference trajectory calculations, and we achieve an improvement of energy conservation  of an order of magnitude for long duration integrations. Using extended precision sparingly like this allows the compiler to optimise the majority of the code to make use of single instruction multiple dispatch (SIMD) operations which are not generally available for extended precision variables on Intel or AMD64 hardware.

Figure~\ref{fig:absolute state variable precision} shows an example of the relative scales of the various state variables used within TES: the state vector variables are in blue and the compensation variables are in orange. The cross-hatched area shows the extra precision required in the reference trajectory, $\qosc$, such that the extra relative precision in $\dq$ can be used to create a scheme with a round-off error, per step, of $10^{-19}$, assuming that $\dq$ remains suitably small. The cross-hatched area is also exactly the extra precision that can be obtained through the use of long doubles in the calculation of the reference trajectories, and, as will be shown in Section \ref{sec: validation of implementation}, allows for a reduction in energy violation of over an order of magnitude.

\begin{figure*}
    \centering
    \includegraphics[width=0.95\textwidth]{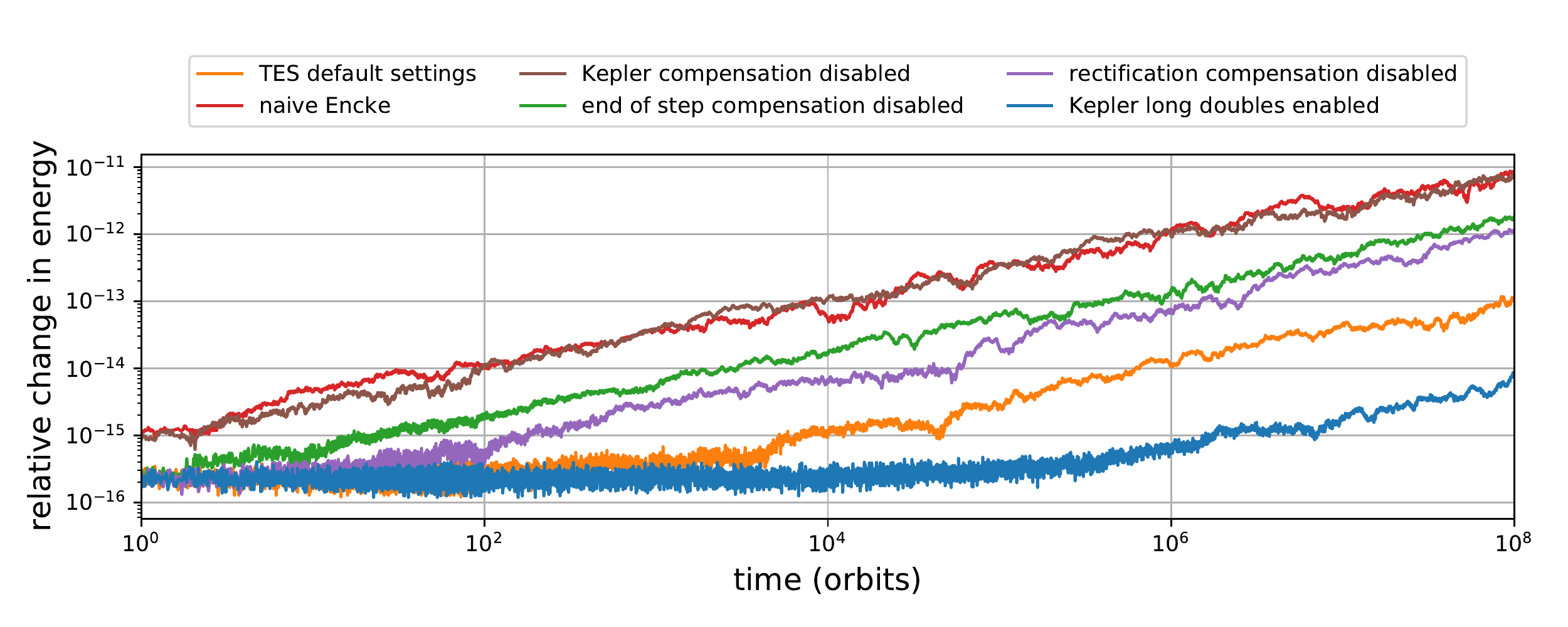}
    \caption{The effect of each numerical implementation technique on the relative change in energy, $dE/E$, for simulations of the inner solar system over $10^8$ Mercury orbits. All results plotted are the RMS of twenty realisations of the initial conditions randomly perturbed on the order of $10^{-15}$. All results use the double precision implementation of TES unless stated otherwise. "TES default settings" has all compensation features enabled, while "naive Encke" has all compensation features disabled.}   
    \label{fig: fancy features energy}
\end{figure*}

\subsection{Numerical Solution}
\label{sec: implementation numerical}
In a similar fashion to Eq.~\eqref{eq: update qosc compensated}, the numerical integrator also obtains a compensation variable at the end of a step for each variable being integrated, as is shown in the top panel of Fig.~\ref{fig: compensation diagram}. In our case, this means that at the end of an integration step we obtain $\{\dq, \dq*\}$ and $\{\dpp, \dpp*\}$. We therefore have two separate sets of compensation variables: one for the reference trajectories, $\qosc*$ and $\posc*$, and one for the deltas, $\dq*$ and $\dpp*$; however, these two sets must be combined at the end of a step to ensure the correct error is used in the subsequent step. Combination of compensation variables is performed at the end of each step as can be seen by label $3$ in Fig.~\ref{fig: compensation diagram}. It is achieved through the following algorithm:
\begin{equation}
    \nonumber
    \begin{split}
            &\boldsymbol{\psi}* \leftarrow 0, \\ 
            &\{\qosc, \, \boldsymbol{\psi}* \} \leftarrow (\{\qosc, \, \boldsymbol{\psi}*\} \ominus \dq*) \ominus \qosc*, \\
            &\dq* \leftarrow 0,  \quad \qosc* \leftarrow 0, \\
            &\{\dq, \, \qosc* \} \leftarrow \{\dq, \, \qosc* \} \ominus \boldsymbol{\psi}*
    \end{split}
\end{equation}
where $\boldsymbol{\psi}*$ is a temporary summation variable for each body.
This algorithm begins by combining the two compensation variables $\qosc*$ and $\dq*$ into a new compensation variable $\psi*$. This is done as a compensated subtraction into the reference trajectory $\qosc$ in case the subtraction causes the range of the double precision variable $\psi*$ to overlap with $\qosc$. After this, the range of $\psi*$ now overlaps with the least significant region of $\dq$ and a third compensated subtraction is therefore used to update $\dq$ accordingly. In this final subtraction the reference trajectory compensation variable, $\qosc*$, is used to store the final summation error such that it can be used immediately in the subsequent Kepler step. The same process is also used for the momenta terms, $\posc$ and $\dpp$. Note that this process differs from rectification, described next, as only the reference trajectories are used here. 

\subsection{Rectification}
\label{sec: implementation rectification}
Compensated summation can also be applied to maintain precision across a rectification by following a very similar algorithm:
\begin{equation}
    \nonumber
    \begin{split}
            &\boldsymbol{\psi}* \leftarrow 0, \\ 
            &\{\qosc, \, \boldsymbol{\psi}* \} \leftarrow \{\qosc, \, \boldsymbol{\psi}*\} \oplus \dq, \\
            &\{\qosc, \, \boldsymbol{\psi}* \} \leftarrow \{\qosc, \, \boldsymbol{\psi}*\} \oplus {\qosc*},  \\
            &\dq \leftarrow -\boldsymbol{\psi}* \\
            &\qosc* \leftarrow 0, \quad \dq* \leftarrow 0
    \end{split}
\end{equation}
where again $\boldsymbol{\psi}*$ is a temporary summation variable for each body. This algorithm begins by performing the rectification process by summing the reference trajectories and deltas together and using temporary summation variables to capture any lost precision from the addition. The reference trajectories, $\qosc$, are then refined using the compensation variables $\qosc*$ and $\dq*$ with any lost precision being captured again by $\boldsymbol{\psi}*$. Due to the sign convention used in the compensated summation this means that $-\boldsymbol{\psi}*$ now contains the rectified value of the deltas which is then placed into $\dq$. This process is also used to rectify the momenta, $\posc$.

\begin{figure*}
    \centering
    \includegraphics[width=0.95\textwidth]{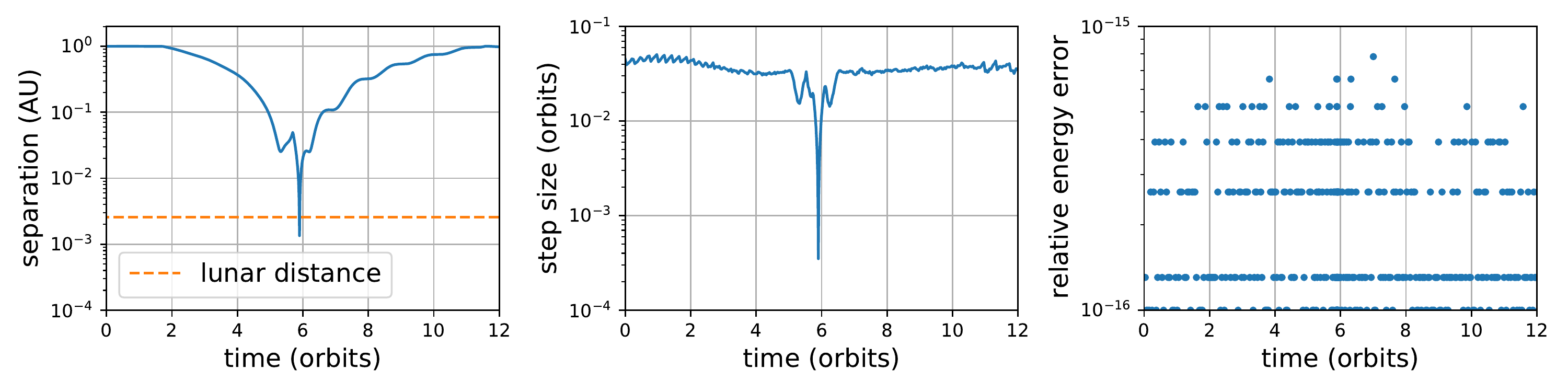}
    \caption{A close encounter between two Earth mass planets orbiting a solar mass star at roughly $1$~AU. The left panel shows the minimum separation between bodies over time. The orange line shows the separation between the Earth and Moon. The central panel shows the step size used by TES during the encounter. The right hand panel shows the relative energy error over the same time span.}   
    \label{fig: close encounter handling}
\end{figure*}

\section{Validation of Implementation Details}
\label{sec: validation of implementation}

In the previous section, we described four key numerical features present in TES, in summary they are:

\begin{enumerate}[leftmargin=*]
        \item Kepler compensation  described in Section~\ref{sec: implementation analytical}.
        \item Kepler long doubles described in Section~\ref{sec: implementation analytical}.   
        \item Combining compensation variables described in Section~\ref{sec: implementation numerical}.
        \item Rectification compensation described in Section~\ref{sec: implementation rectification}.
\end{enumerate}
 With the exception of Kepler long doubles, the default configuration of TES enables all of these features. Figure~\ref{fig: fancy features energy} shows the effect of disabling compensation at various points in TES has upon the energy conservation in a simulation of the inner planets of our solar system over a period of $10^8$ Mercury orbits. The data plotted is the RMS of twenty realisations of the initial conditions randomly perturbed on the order of $10^{-15}$. The modification to the force function in Section~\ref{sec: implementation DHEM} to avoid numerical issues due to cancellation of similar size terms is enabled for all configurations as without it a step size lockup can occur. 

As a baseline for the performance without any numerical improvements a \emph{naive Encke} implementation is included that has no numerical improvements other than the reformulation of the acceleration. The default configuration of TES using only double precision with all compensated summation schemes enabled is shown under \emph{TES default settings}. Here, the conservation of energy for TES is two orders of magnitude better than for the naive Encke scheme. If the use of extended precision floating point variables is permitted in the Kepler solver then the energy conservation in TES can be further improved by an order of magnitude as compared to the default configuration. Disabling individual compensation schemes results in energy conservation at least ten times worse than that of TES with default settings. In the worst case, when compensated summation is not used in the final update step of the $f$ and $g$ functions, the conservation in energy can be as poor as just using the naive Encke method which highlights how important a precise solution to the dominant Keplerian motion is to schemes of this nature.

In order to ensure that TES can handle close encounters we ran a simulation of three Earth mass planets orbiting a solar mass star at $1$~AU as per \citet{Bartram2021}. Planets are tightly packed and over time the system becomes unstable causing close encounters between them. Figure~\ref{fig: close encounter handling} shows one of these encounters where the planets passed closer to one another than the Moon is to the Earth. Firstly, in the left hand panel, a close encounter after approximately five orbits leads to a much closer encounter an orbit later. Next, in the central panel the step size controller shrinks and subsequently expands the step size appropriately. Finally, in the right hand panel, the relative energy error can be seen to be maintained below $10^{-15}$ throughout. Interestingly, the relative energy error values in this plot take six distinct values indicating that the error is confined to the three least significant bits of our double precision representation of the energy. 

These examples validate that the TES model derived in Section~\ref{sec:TES Model} and implemented as described in Section~\ref{sec: implementation details} is indeed capable of performing highly accurate long term integration as well as handle close encounters between bodies effectively.

\section{Numerical Experiments}
\label{sec: numerical experiments}

To further investigate the performance of TES in a variety of settings, in this section we perform a series of numerical experiments. We also provide comparisons against a number of other integrators. The schemes used are: TES (double); TES (long double), which makes use of long doubles in the Kepler solver; naive Encke; ias15 \citep{Rein2014} from the Rebound package \citep{Rein2012}; and Bulirsch-Stoer \citep{Bulirsch1966} as well as the hybrid \citep{Chambers1999} integrators from the MERCURY package. Table~\ref{tab: integrator tolerances} contains the default tolerances used throughout these experiments unless otherwise specified. In the case of the TES, naive Encke and ias15 schemes the tolerances used are the recommended defaults. All runtime measurements were performed on an Intel Core i7-6700 CPU running at 3.4 GHz.

\begin{table}
	\centering
	\caption{Summary of all default integrator tolerances used. TES, naive Encke and ias15 tolerances are the recommended defaults.}
	\label{tab: integrator tolerances}
	\begin{tabular}{lcc} 
		\hline
		Tool & Default Tolerance \\
         \hline
        TES (double)      &   $10^{-6}$   \\
        TES (long double) &   $10^{-6}$   \\
        naive Encke       &   $10^{-6}$   \\
        ias15             &   $10^{-9}$   \\
        Bulirsch-Stoer    &   $10^{-14}$  \\
        hybrid            &   $10^{-14}$ with 20 steps per orbit  \\
        \hline
	\end{tabular}
\end{table}

\subsection{Efficiency Mass Dependence}
\label{sec: mass dependence}

\begin{figure}
    \centering
    \includegraphics[width=0.475\textwidth]{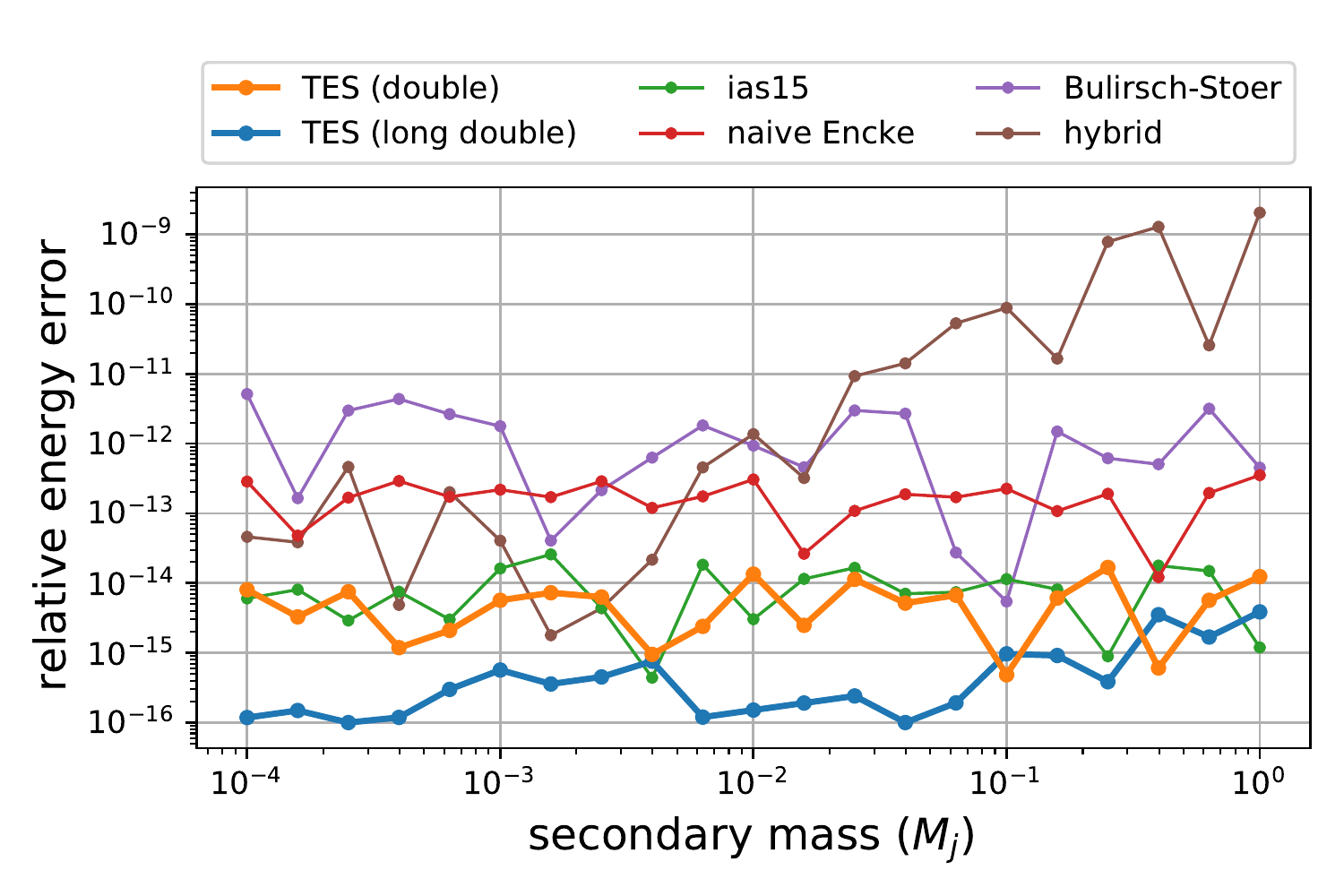}
    \caption{Relative energy error of simulations of circular two-body systems over $10^4$ orbits for all integrators. The primary is a solar mass star and the mass of the secondary is varied across a range coincident with that of our solar system. The Encke based methods must account for the motion of the central body and the two-body problem is therefore still an appropriate test case.}
    \label{fig: mass de}
\end{figure}

\begin{figure}
    \centering
    \includegraphics[width=0.475\textwidth]{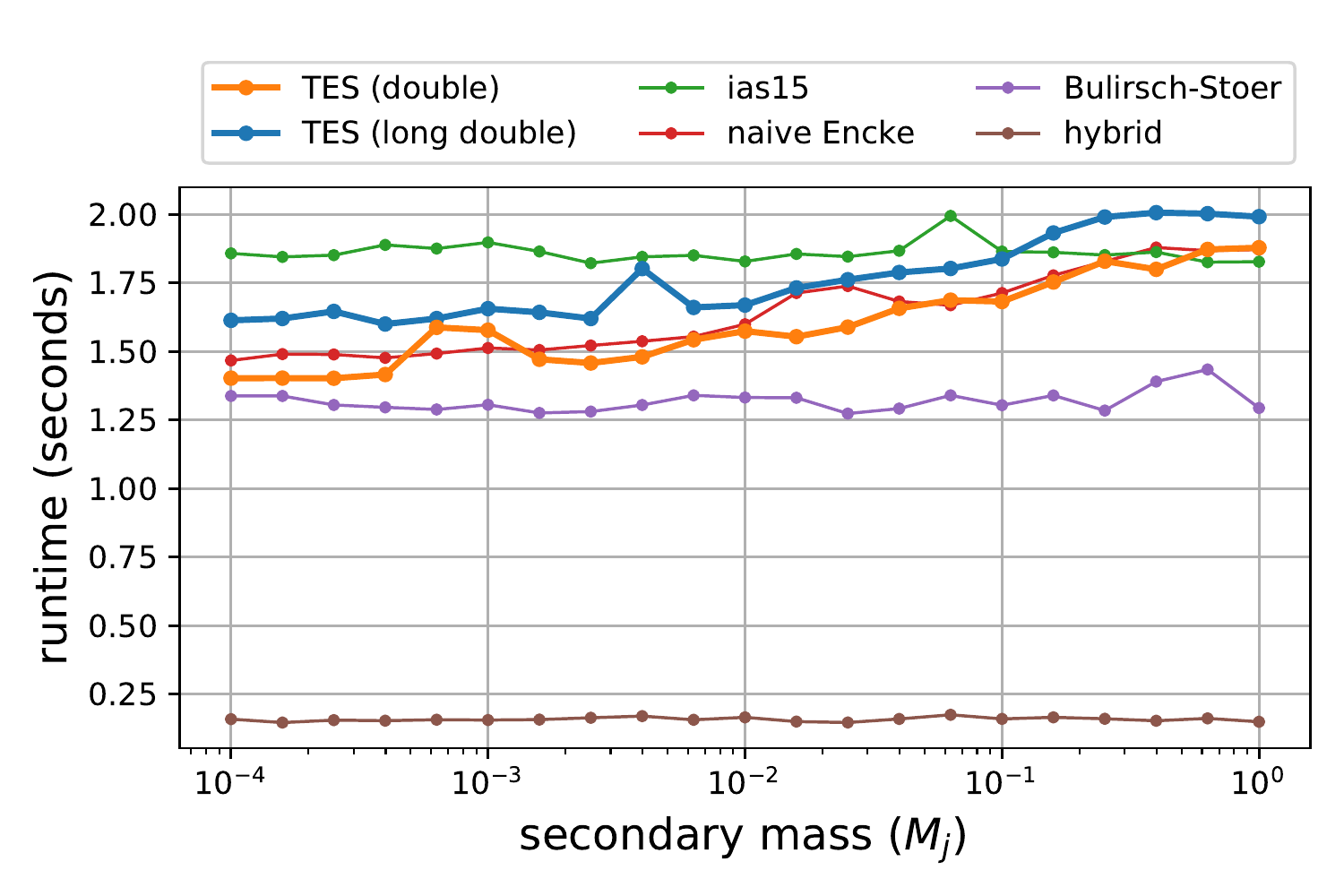}
    \caption{Runtime of simulations of two-body systems over $10^4$ orbits for all integrators. The primary is a solar mass star and the mass of the secondary is varied across a range coincident with that of our solar system. The Encke based methods must account for the motion of the central body and the two-body problem is therefore still an appropriate test case. Each data point is the average of twenty identical integrations.}   
    \label{fig: mass rt}
\end{figure}

As discussed previously, the magnitude of the deltas in comparison to the magnitude of the reference trajectory, the delta ratio, must be kept small in order to maximize the efficiency of an Encke method. Simultaneously, one must not rectify too frequently as rectifications degrade the precision of the predictor in the subsequent step and thus incur a performance penalty. In the absence of close encounters, the dominant contribution to the acceleration of the deltas is related to the motion of the central body which in turn is dependent on the system mass ratio. Therefore, this experiment is designed to understand in which regions of system mass ratio TES is effective.

We perform integrations of twenty-one two-body problems for our full selection of integration packages. We have opted to examine the range of system mass ratios that can be found in our own solar system if each planet is taken in isolation with the Sun. Therefore, this experiment ranges from a secondary of Jupiter mass, $M_j$, down to a mass of $10^{-4} M_j$, approximately equal to that of Mercury. The samples across the range of masses are logarithmically spaced, and the primary is always a solar mass star. The secondary body is placed on a circular, co-planar orbit at $1$~AU and integrations are performed for $10^4$~orbits. Runtime is calculated as the mean of twenty identical integrations for each two-body problem for each integrator.

Figure~\ref{fig: mass de} shows the relative energy error achieved in these experiments. Here, with the exception of the hybrid scheme, we find no dependence between the relative energy error and the system mass ratio, simply meaning that the error control algorithms within each integrator are performing as expected. However, we do find large differences in the precision of the various integration schemes. In particular, the Bulirsch-Stoer, hybrid and naive Encke schemes all fail to reach the regions of highest energy conservation. In contrast, the schemes that are floating point arithmetic aware, i.e TES and ias15, are much more precise. TES (double) and ias15 have almost identical performance across the entire range of system mass ratios examined. TES (long double) is the best performer overall and outperforms TES (double) and ias15 by up to two orders of magnitude.

Figure~\ref{fig: mass rt} shows the runtime for the same experiments where a cluster of curves can be seen as well as the hybrid scheme which is between six and eight times faster than the non-symplectic schemes. We find that the standard deviation in runtime across all TES realisations is $146$~ms. The Bulirsch-Stoer, hybrid and ias15 schemes can be seen to exhibit no dependency of the runtime on the system mass ratio. However, all of the Encke based schemes, i.e. TES and naive Encke, show a positive correlation between the runtime and the system mass ratio, as predicted. An interesting comparison is that of TES (double) and ias15 where for systems with a smaller mass ratio TES is able to achieve the same levels of precision with only $75\%$ of the computational cost. TES (double) remains more efficient until the mass of the secondary is roughly $10^{-1} M_j$. Therefore, for maximum benefit, TES should be applied to systems with a system mass ratio below this value. Finally, TES (long double) can also be seen to perform well with a runtime below ias15 despite having a better conservation in energy of up to two orders of magnitude.

\begin{figure}
    \centering
    \includegraphics[width=0.475\textwidth]{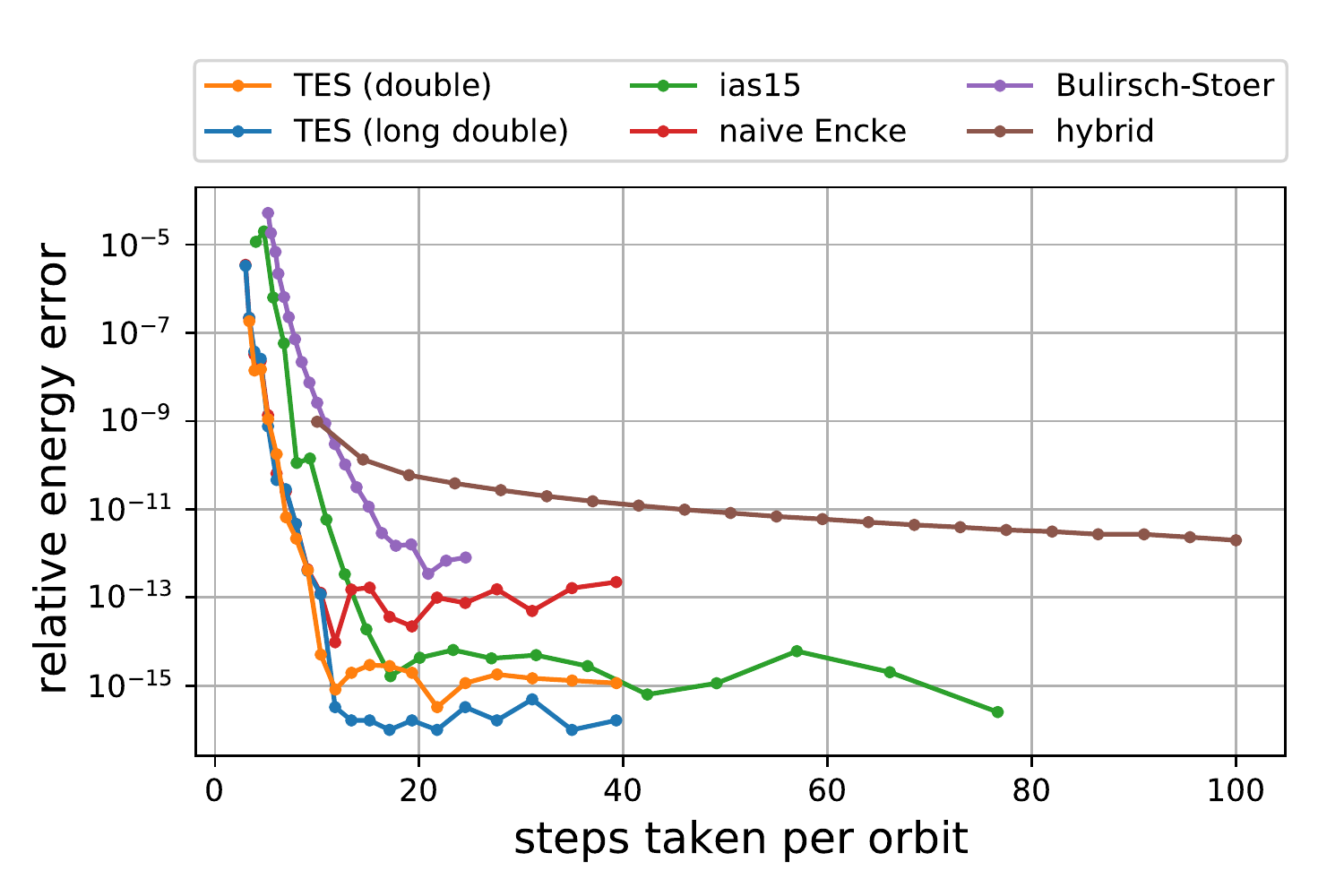}
    \caption{Relative energy error against average number of steps per orbit for the inner solar system for $10^4$ Mercury orbits.} 
    \label{fig: tolerance steps taken}
\end{figure}    

\begin{figure}
    \centering
    \includegraphics[width=0.475\textwidth]{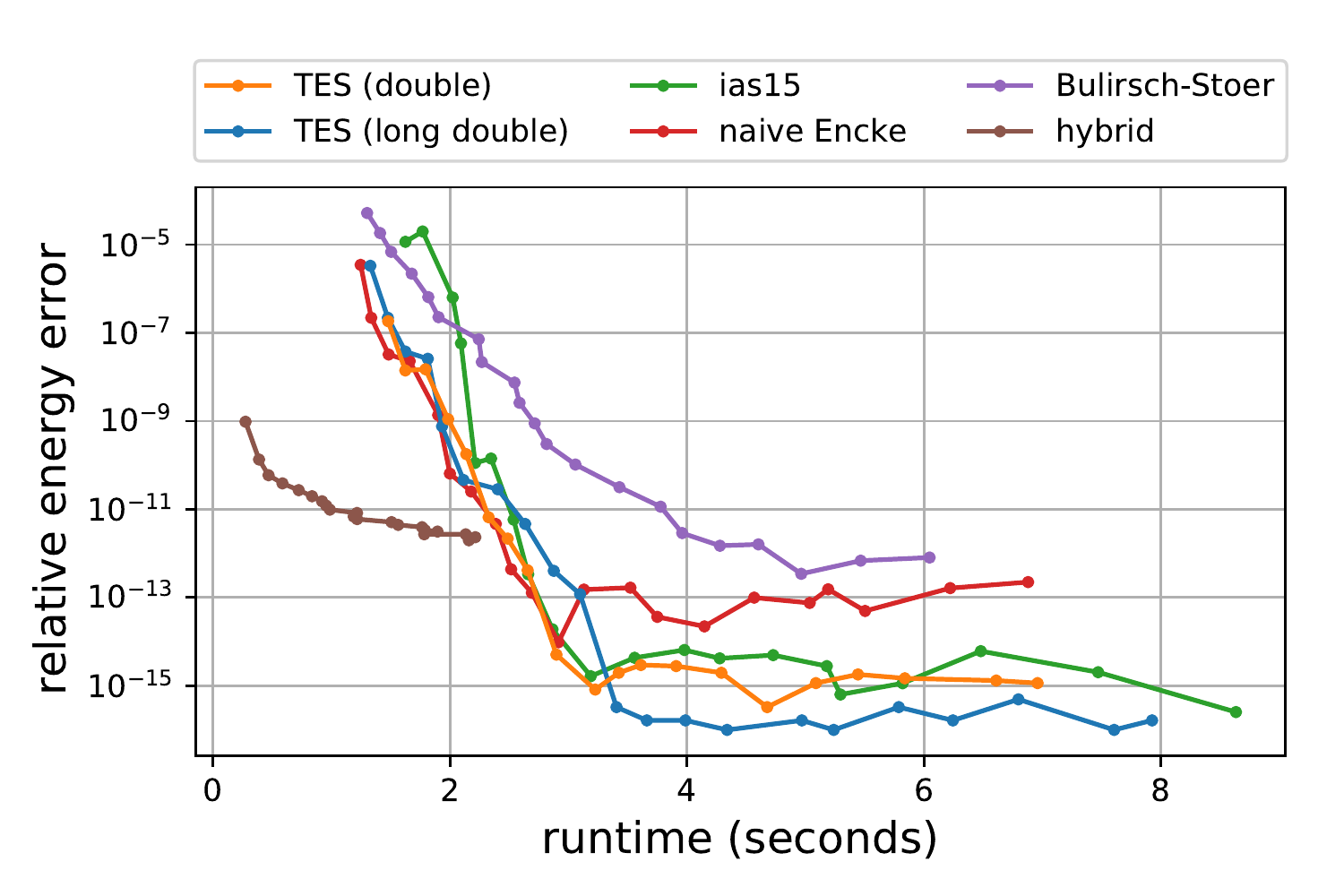}
    \caption{Relative energy error against runtime for the inner solar system for $10^4$ Mercury orbits.}   
    \label{fig:runtime de}
\end{figure}

\subsection{Convergence and Runtime Comparisons}
\label{sec: convergence and runtime comparisons}
Next, we study the convergence and runtime of TES in comparison to the wider field of integrators over a period of ten thousand orbits of the innermost planet. In the previous section we showed that TES is most effective in systems with a low system mass ratio, and we have therefore chosen the inner planets of our solar system as a test problem using initial conditions taken from the NASA Horizons database \citep{NASA2021}. The tolerance of each of the non-symplectic integrators is varied over a range of values such that the relative energy error no longer converges. TES and ias15 use a range ending at the recommended operating tolerances in Table~\ref{tab: integrator tolerances}. The hybrid scheme uses a fixed tolerance of $10^{-14}$ throughout but the step size, which must be kept constant within an integration, is varied until the relative energy error no longer converges.

Figure~\ref{fig: tolerance steps taken} shows the relative energy error against the number of steps per orbit. Once again we can see a divide between the optimal and non optimal schemes, TES and ias15 clearly conserve energy more precisely than the other schemes with TES (long double) being the most precise by roughly an order of magnitude once the round-off error dominated regime is entered at roughly fifteen steps per orbit. The power of the Encke method can be seen in two places in this plot. Firstly, in the truncation error dominated region, i.e. the region below roughly fifteen steps, where the relative change in energy for a given step size is approximately three orders of magnitude smaller than any of the direct integrations. Secondly, in the furthest right data point for TES and ias15 where the recommended default tolerances for the Encke based methods yield a reduction in the number of steps taken per orbit when compared to a direct integration. Figure~\ref{fig:runtime de} shows how these benefits manifest themselves in the runtime. Immediately, the hybrid scheme can be seen to stand apart from the others and is indeed much faster than any of the non-symplectic integrators; however, as we will show in Section~\ref{sec: numerical experiments apophis} the relatively low precision of the hybrid scheme is not entirely suitable to modelling exoplanet evolution in the presence of close encounters. The Bulirsch-Stoer scheme has a poor runtime in comparison to the other integrators and does not reach the highest levels of precision either. The naive Encke method has a favourable runtime in the truncation error dominated regime but is not capable of the energy conservation of the optimal, floating point implementation aware methods. Of the three remaining integrators, TES (double), TES (long double) and ias15, the performance is similar. However, for the recommended default tolerances, the furthest right data point for each integrator, TES (double) is the fastest and is approximately $20\%$ faster than the slowest scheme. Interestingly, TES (long double) has the best energy conservation by up to an order of magnitude and has very comparable runtime to ias15 despite the disadvantage of non being able to use vectorisation in the Kepler solver.

\subsection{Long-term Integrations of the Inner Solar System}
\label{sec: numerical experiment longterm integrations}

    \begin{figure*}
        \centering
        \includegraphics[width=\textwidth]{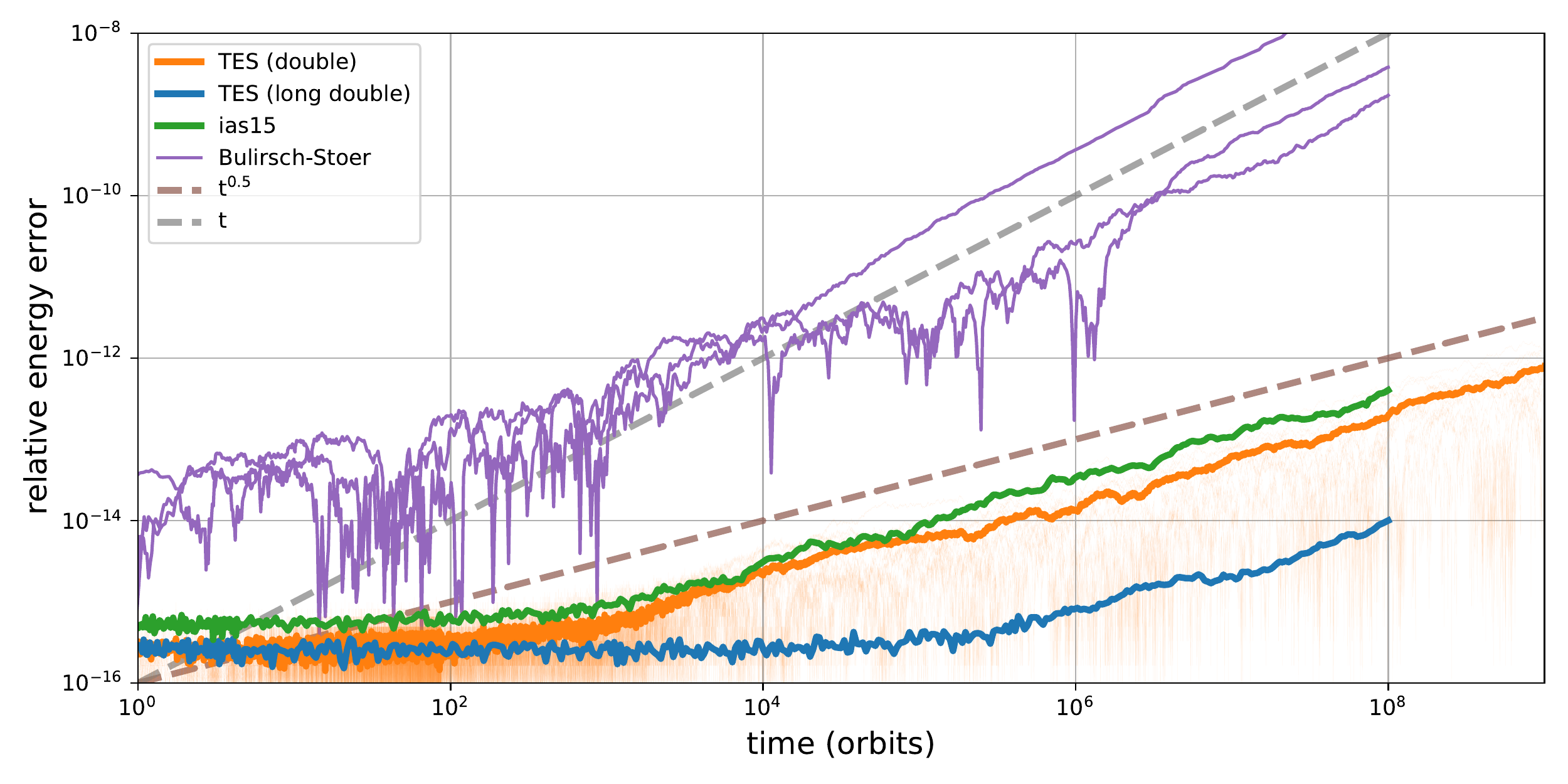}
        \caption{Relative energy error of long-term simulations of the inner solar system lasting either $10^8$ or $10^9$ Mercury orbits using default tolerances for TES and ias15. Bulirsch-Stoer is included for comparison with manually chosen tolerances of $10^{-13,-14,-15}$ to maximise precision. For TES and ias15 lines plotted are the RMS of twenty realisations of the initial conditions perturbed on the order of $10^{-15}$. Individual realisations are also shown for the TES (double) integrator. Slopes show optimal ($\sqrt{t}$) and linear error growth in brown and grey, respectively.}
        \label{fig: long term conservation}
    \end{figure*}        

In the field of exoplanet modelling it is typical for simulations to span a billion dynamical periods. It is therefore of great importance to ensure that propagation schemes follow the optimal error growth of Brouwer's Law, i.e. $\propto \sqrt{N}$ for the relative energy error, where $N$ is the number of steps taken. We have therefore chosen to perform long-term simulations of the inner solar system to ensure TES conforms to this requirement. We use the tolerance in Table~\ref{tab: integrator tolerances} for TES and ias15. In keeping with the original ias15 experiments \citep{Rein2014} and to generate a statistical sample we perform twenty integrations for TES and ias15 with a perturbation in the initial conditions on the order of $10^{-15}$. The RMS of these twenty realisations is plotted in Fig.~\ref{fig: long term conservation}. Additionally, results of three integrations highlighting the performance of the Bulirsch-Stoer integrator are also shown for tolerances of $10^{-13}$, $10^{-14}$ and $10^{-15}$. Integrations performed with TES (double) span the full $10^9$ Mercury orbital periods whereas all other schemes are terminated after $10^8$ Mercury orbital periods to save on computation. Finally, two slopes are included: one in grey marking the linear error growth typically associated with truncation dominated regimes, and another, in brown, showing the optimal error growth associated with the symmetrical distribution of round-off error required for Brouwer's Law.

Figure~\ref{fig: long term conservation} highlights that TES (double) and TES (long double) both follow Brouwer's law for the full integration duration and therefore show that these schemes are well suited to the long duration integrations required in exoplanet modelling. Note that despite the larger steps taken by TES (double) the use of the Encke method has enabled the truncation error growth to be suppressed for the entirety of integrations. When performed by an integrator that follows Brouwer's law, the RMS relative energy error of a suitably large number of realisations $\epsilon \approx C \sqrt{N}$ where $C$ is a constant approximately at floating point precision, i.e. $\approx 10^{-16}$, and $N$ is the number of steps taken. In double precision, TES performs marginally better than ias15 and we believe this is due to the larger step size taken by the Encke method reducing the $\sqrt{N}$ term. TES (long double) performs only the solution of the Keplerian motion in extended precision, i.e. the integrator and force models use only double precision. However, even with this sparing use of extended precision TES is able to attain a relative energy error of over an order of magnitude better than if only double precision is used throughout, and importantly, it does this without excessive computational cost. Most notably, TES (long double) is able to integrate for up to $10^5$ orbits before there is any noticeable growth in relative energy error above the floating point floor. Both TES (double) and ias15 already start to show error growth after just hundreds of orbits.

\begin{figure}
    \centering
    \includegraphics[width=0.475\textwidth]{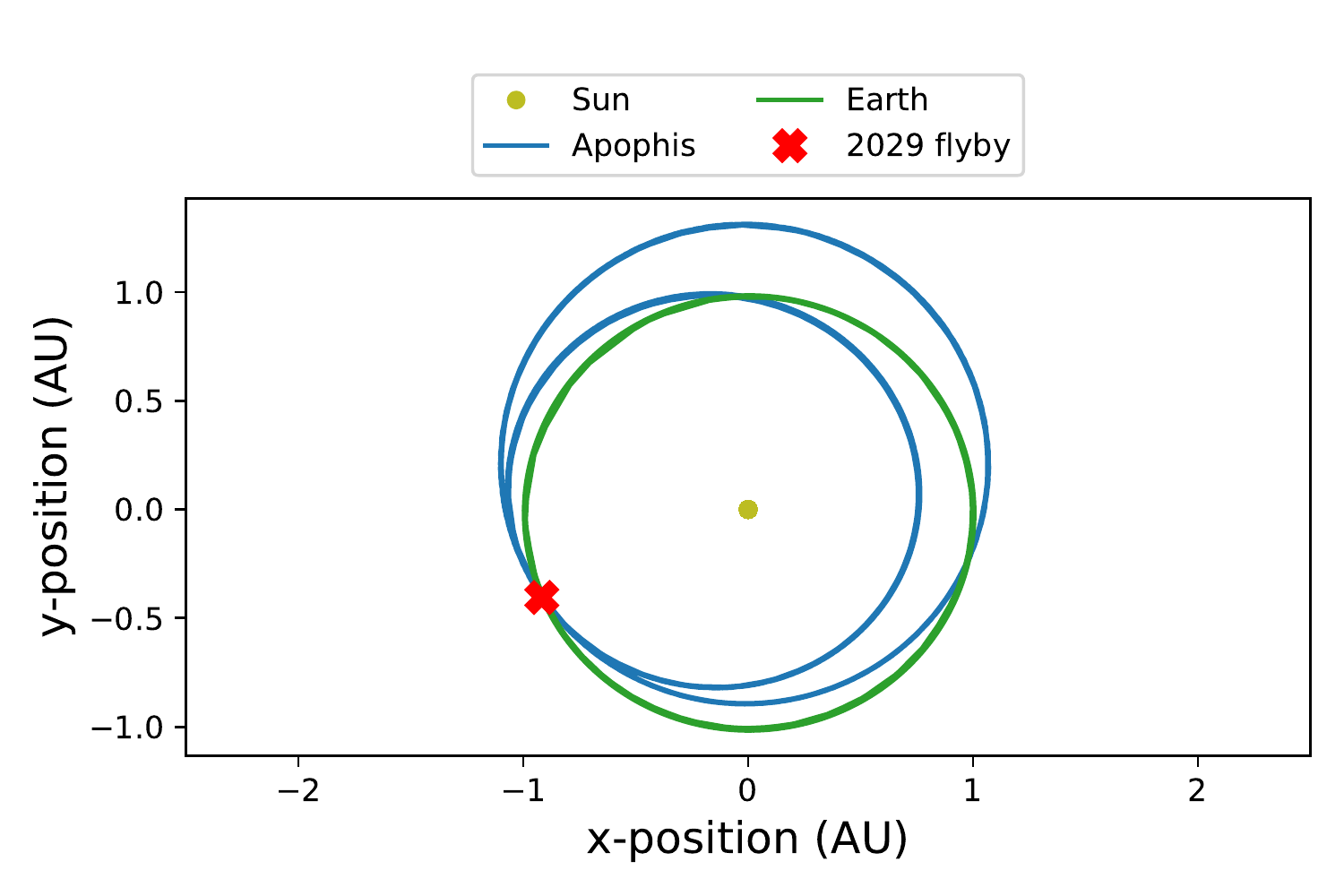}
    \caption{Orbits of the Sun, Earth and Apophis over a one-hundred year period from $1979$ to $2079$. The closest approach is marked and causes a transition of Apophis from Apollo to Atens group.}
    \label{fig:apophis orbital diagram}
\end{figure}

\begin{figure}
    \centering 
    \includegraphics[width=0.475\textwidth]{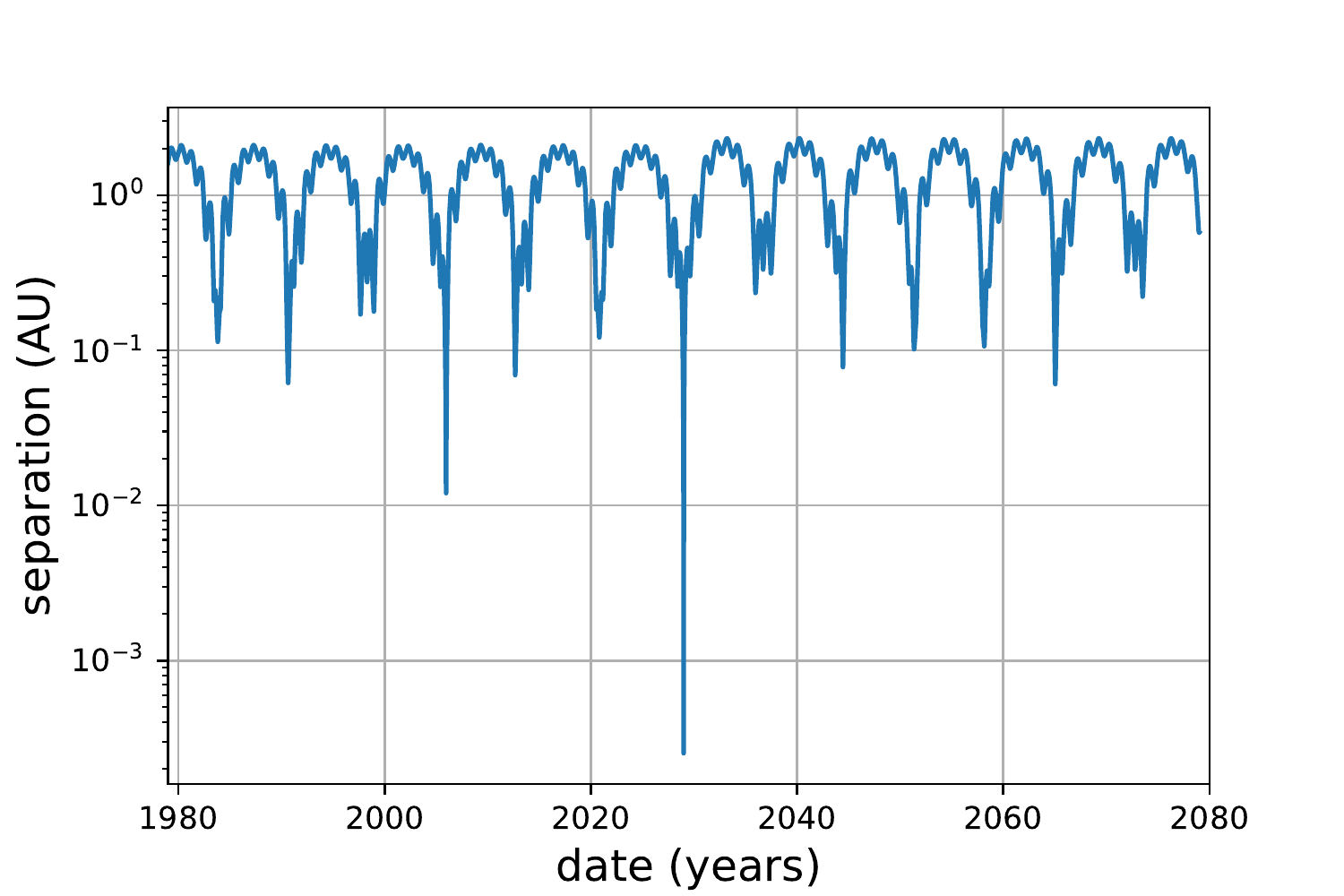}
    \caption{Relative separation between the Earth and Apophis over a one hundred year period from $1979$ to $2079$. The closest approach is approximately $2.5 \times 10^{-4}$~AU or roughly $17,000$~km, well within the geosynchronous orbital altitude at $35,786$~km.}
    \label{fig: apophis separation}
\end{figure}

\begin{figure}
    \centering
    \includegraphics[width=0.475\textwidth]{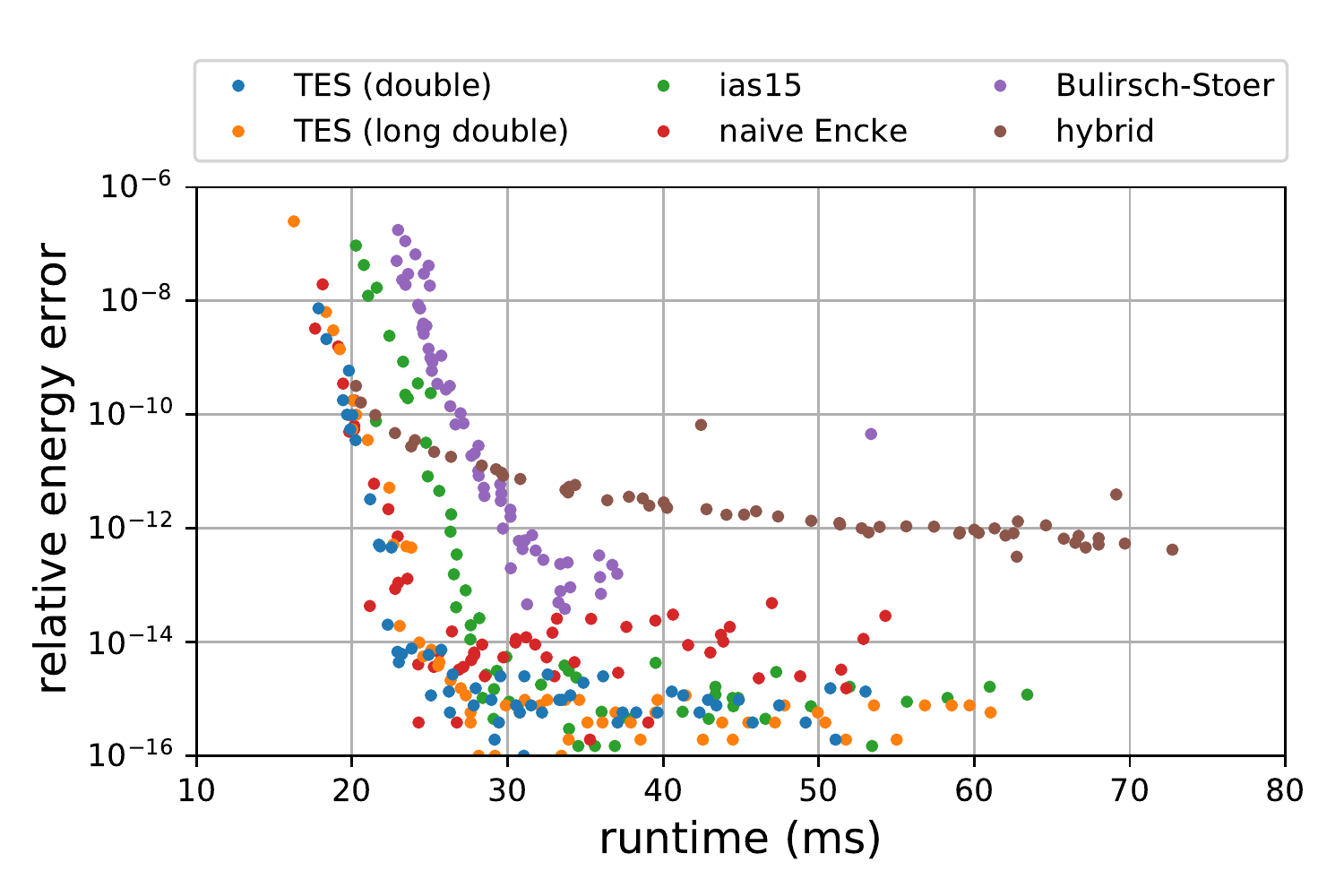}
    \caption{Relative energy error for a given runtime at the end of a one hundred year integration of the Sun, Earth and Apophis, including the 2029 close encounter with Earth. Each data point is the mean of twenty identical integrations.}   
    \label{fig: apophis de vs rt}
\end{figure}    

\begin{figure}
    \centering
    \includegraphics[width=0.475\textwidth]{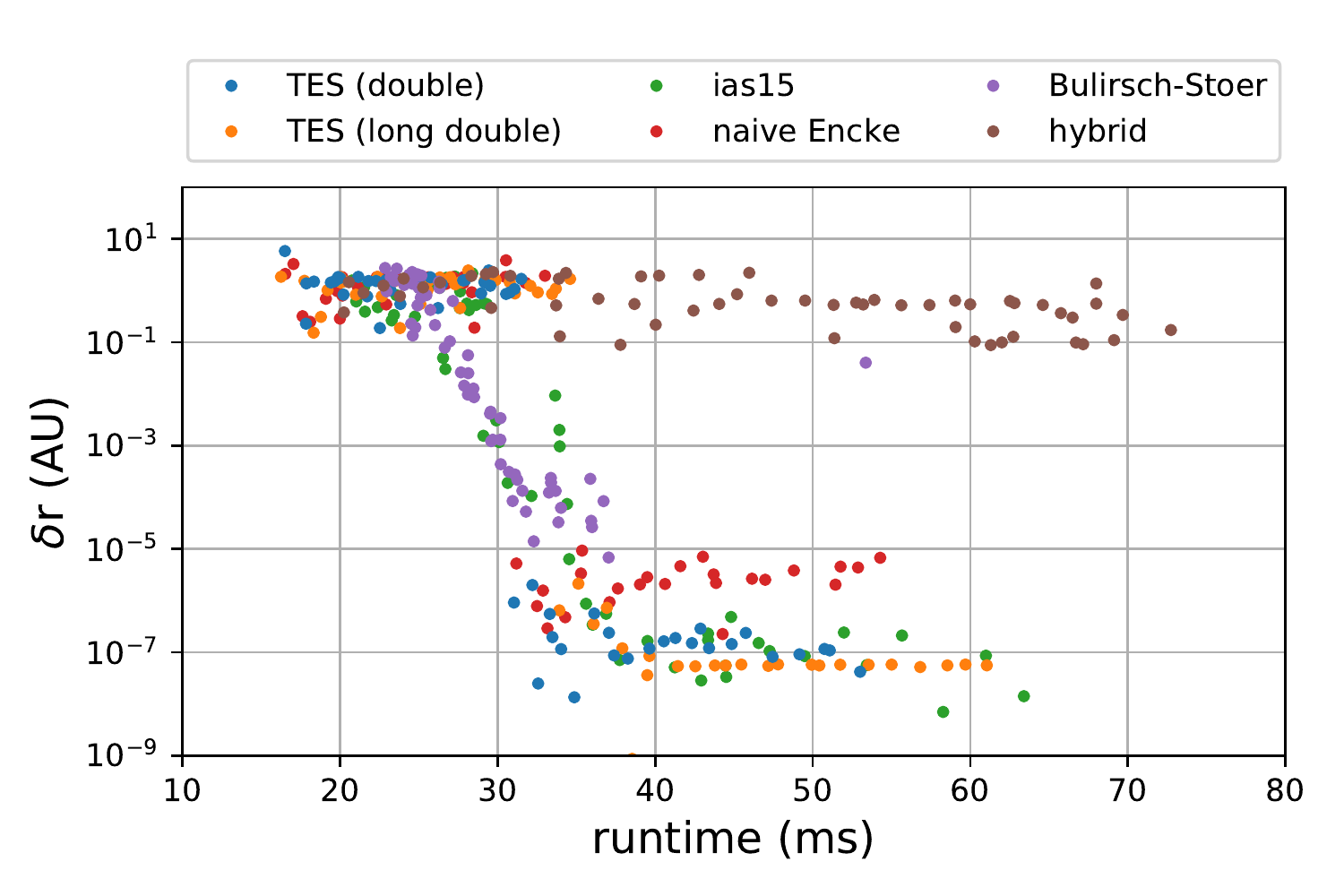}
    \caption{Final position error for a given runtime of Apophis after a one hundred year integration of the Sun, Earth and Apophis, including the 2029 close encounter with Earth. Each data point is the mean of twenty identical integrations.}   
    \label{fig: apophis dr vs rt}
\end{figure}    

\subsection{Apophis 2029 Encounter}
\label{sec: numerical experiments apophis}
The last example we present studies the performance of TES in the presence of close encounters. When the dynamics of planetary systems are such that bodies do not remain well separated indefinitely it is important that close encounters are handled accurately, e.g. when modelling the behaviour of systems after an instability event \citep{Rice2018, Bartram2021}. Closer to home, another example is the accurate modelling of near Earth objects (NEOs) to understand potential hazards to humanity. Of the known NEOs, one of particular interest is Apophis owing to the fact that it will pass within $100,000$~km of the Earth in $2029$ bringing it closer than the Moon.

The simplified model we use \citep{Amato2017} evolves only the Sun, Earth, and Apophis and makes use of purely Newtonian dynamics throughout. While not a realistic model, it allows us to showcase the behaviour of different integrators in the presence of strongly perturbing close encounters. To obtain a set of initial conditions and a high precision reference solution trajectory, first the position of the three bodies were obtained from the NASA Horizons system at the point of closest approach during the $2029$ encounter. Next, a Dormand-Prince integrator \citep{Prince1980} operating in quadruple precision was used to integrate backwards for fifty years to obtain the initial conditions. Finally, the same integrator was used to propagate the initial conditions forward for one hundred years to obtain the final conditions. We consider this high precision integration the truth, such that the final positions can also be used as an error metric in addition to energy conservation.

The trajectories followed by the three bodies are plotted in Fig.~\ref{fig:apophis orbital diagram} and the transition of Apophis from the Apollo to the Atens group after its flyby with Earth can clearly be seen. Figure~\ref{fig: apophis separation} shows the separation between Apophis and the Earth over the simulated period of $100$ years. Here, we see that Apophis makes many close approaches within 10 Earth Hill radii, or $0.1$~AU, but only one very close approach which is in $2029$. This approach distance is approximately $2.5 \times 10^{-4}$~AU or roughly $17,000$~km, well within the geosynchronous orbital altitude at $35,786$~km.
We perform integrations at 61 different tolerance settings for each of the integrators in Table~\ref{tab: integrator tolerances}. The tolerances used are chosen in the same way as in Section~\ref{sec: convergence and runtime comparisons}. At each tolerance we perform twenty identical integrations and take the mean value for the runtime.

Figure~\ref{fig: apophis de vs rt} shows the relative energy error for a given runtime. First, consider the behaviour of TES (double) in this plot. We find that in the truncation dominated regime it is amongst the fastest schemes competing only with the naive Encke and TES (long double). TES (double) can also be seen to be fastest at the default tolerance (Table~\ref{tab: integrator tolerances}) which is represented by the furthest right blue data point; this is the recommended default setting for users. Once TES (double) has converged such that only round-off error is present it is clear that the numerical implementation described in Section~\ref{sec: implementation details} has improved the performance by between one and two orders of magnitude in comparison to a naive Encke method (red). Due to the very short integration timescale of only one hundred orbits TES (long double) shows no advantage over TES (double) and the runtime for the default tolerance setting is comparable to that of ias15.
The MERCURY integrators, Bulirsch-Stoer and hybrid, fail to conserve energy as precisely as the schemes based upon Everhart's Radau and reach final values of relative energy error of $10^{-13}$ and $10^{-12}$, respectively. Interestingly, in contrast to Fig.~\ref{fig:runtime de}, we can see that in the presence of repeated close encounters the hybrid scheme runtime increases to be comparable to that of the non-symplectic integrators.

While the conservation of energy is a common metric for the  accuracy, its value is usually dominated by the most massive objects in a given system. Given the low mass of Apophis in comparison to the Earth we also look at the final position of Apophis as output by each integrator in comparison to our quadruple precision numerical solution. Figure~\ref{fig: apophis dr vs rt} shows the error in the position of Apophis, $\delta r$, for a given runtime. Again starting with TES (double), there are two regions of performance: one where the runtime is low, below $32$~ms, and it performs poorly, and one where the runtime is higher, above $32$~ms, where the scheme performs well. What is interesting is the lack of a transition period between these two regions which is present in, e.g., ias15; instead, all of the Encke based methods are either highly precise or highly inaccurate. Fortunately, the recommended default tolerance for both implementations of TES is well within the highly precise solution region. Given a suitable tolerance we can see that both implementations of TES and ias15 are comparable in precision, with TES being slightly faster for the default tolerances (Table~\ref{tab: integrator tolerances}).

For this problem, TES (long double) shows interesting behaviour once it has converged to roughly $\delta r = 10^{-7}$~AU. At this point, its performance becomes highly consistent regardless of decreases in tolerance. This is in contrast to all of the other Everhart based schemes which exhibit variance in the precision of the final position of Apophis with changing tolerance. We find that the precisions seen here are highly consistent with \citet{Amato2017} who find their implementation of Everhart's Radau scheme integrating using Cowell's formulation converging to roughly $10^{-7}$~AU. In order to obtain more precise solutions authors had to employ regularisation techniques on the equations of motion.

The positional precision obtained after a close encounter is highly important if one wishes to model repeated encounters. Deviations from a true trajectory are greatly amplified during a close encounter and any inaccuracies in modelling the first encounter in a series are therefore increased in all subsequent encounters. We find that the precision achieved by TES in these experiments equates to a positional error at the time of closest approach of approximately $10$~cm whereas the naive Encke method achieves an error of $10$~m. Given that the size of keyholes, i.e. regions of space on the b-plane formed by the separation vector at the point of closest approach, found by \citet{Farnocchia2013} that can lead to a resonant return trajectory for Apophis are between $6$~cm and $600$~m, the precision in the naive Encke makes it unsuitable for use in this challenging application domain. Therefore, the numerical improvements in Section~\ref{sec: implementation details} that comprise TES are sufficient to improve a naive Encke method to the point where it can now be used in the study of NEO asteroid dynamics.

We find that the hybrid scheme is not precise enough at the step sizes presented to accurately model the trajectory of Apophis during its flyby with Earth. However, we mirror the findings of \citet{Amato2017} and find that if the hybrid scheme step size is reduced by roughly a factor of one hundred then it is possible to obtain positional errors as low as $10^{-5}$~AU although the computational cost for doing this means this option is of little practical importance. 

\section{Conclusions}
\label{sec: conclusions}

We introduced TES, a new integrator for planetary systems that follows Brouwer's law and permits close encounters between massive bodies. TES builds upon the classical Encke method and takes advantage of the dominant nature of the star in planetary systems. We show that TES is effective across a wide range of planet to star mass ratios but find that the more dominant the central body the more effective the scheme is, with excellent improvements in speed being seen in simulations of the inner solar system.

In Section~\ref{sec: DHEM} we derived a new version of the Encke method in democratic heliocentric coordinates (ENCODE) and presented the equations of motion in this coordinate system. Additionally, we implemented a series of numerical improvements that reduced the round-off error by two orders of magnitude as compared to a naive Encke method. TES is optimal in that it follows Brouwer's law and has an RMS energy error slightly below that of ias15.

We performed extensive comparisons against ias15 in Rebound, and the Bulirsch-Stoer and symplectic hybrid schemes within the MERCURY package. We found that for well separated systems TES is the fastest non-symplectic scheme for a given precision. In the presence of close encounters, we found that TES is able to reach precision much greater than either of the MERCURY schemes, in terms of both conservation of energy and final position, with a precision comparable to ias15.

TES is open source and accessible at \url{https://github.com/PeterBartram/TES}. It is available in a ``double'' version using only double precision floating point arithmetic and a ``long'' version using extended, 80 bit, floating point arithmetic in the Kepler solver. In double precision, we found that TES is only $15\%$ faster than the extended precision implementation which is two orders of magnitude more precise, but for portability we recommend the double precision version as the default. 
Regardless of the version used, we found that TES is faster than ias15 for the same energy conservation in the problems examined although some of this performance may be lost for systems with more massive secondaries. We also found that TES can handle close encounters such as the Apophis flyby in $2029$ efficiently and accurately.

\section*{Acknowledgements}
The authors acknowledge the funding provided by the Engineering and Physical Sciences Research Council (EPSRC) Centre for Doctoral Training in Next Generation Computational Modelling grant EP/L015382/1 that has made this research possible.
Additionally, we acknowledge the use of the IRIDIS High Performance Computing Facility, and associated support services, at the University of Southampton.

\section*{Data Availability}
TES is open source and is available at \url{https://github.com/PeterBartram/TES}.




\bibliographystyle{mnras}
\bibliography{references} 







\bsp	
\label{lastpage}
\end{document}